\documentclass[superscriptaddress,aps,pra,english,floatfix,twocolumn,10pt]{revtex4-1}
\usepackage{graphicx}
\usepackage{float}
\usepackage{natbib}
\usepackage{amssymb}
\usepackage{amsmath}
\usepackage{mathtools}
\usepackage[utf8]{inputenc} 
\usepackage{braket}
\usepackage{graphicx}
\usepackage{subfigure}
\usepackage{pgfplots}
\usepackage{csquotes}
\usepackage{hhline}
\usepackage{amssymb}

\usepackage{dcolumn}
\usepackage{tabularx}
\usepackage[colorlinks=true,linkcolor=blue,citecolor=blue,urlcolor=blue]{hyperref}
\usepackage{longtable}
\usepackage{braket}
\usepackage{float}
\usepackage{listings}
\usepackage{color}
\definecolor{mygreen}{rgb}{0,0.6,0}
\definecolor{mygray}{rgb}{0.5,0.5,0.5}
\definecolor{mymauve}{rgb}{0.58,0,0.82}
\newcolumntype{C}{>{\centering\arraybackslash}X}

\begin{document}
	
	\title{\textbf{Skin effect and dynamical delocalization in non-Hermitian quasicrystals with spin-orbit interaction}}
	\author{Aditi Chakrabarty}
	\email{aditichakrabarty030@gmail.com}
	\affiliation{Department of Physics and Astronomy, National Institute of Technology, Rourkela, Odisha-769008, India}
	\author{Sanjoy Datta}
	\email{dattas@nitrkl.ac.in}
	\affiliation{Department of Physics and Astronomy, National Institute of Technology, Rourkela, Odisha-769008, India}
	\affiliation{Center for Nanomaterials, National Institute of Technology, Rourkela, Odisha-769008, India}
	\date{\today}
	
	\begin{abstract}
	The investigations of the spectral and dynamical delocalization-localization (DL) transition have revealed intriguing features in a wide range of non-Hermitian systems.
 	The present study aims at exploring the spectral and dynamical properties in a non-Hermitian quasiperiodic system with asymmetric hopping in the presence of Rashba 
 	Spin-Orbit (RSO) interaction. In particular, in such systems, we have identified that the DL transition is associated with a concurrent change in the energy spectrum,
 	where the eigenstates always break the time-reversal symmetry for all strenghts of the quasiperiodic potential, contrary to the systems without RSO interaction.
 	Remarkably, we find that the reality of energy spectrum under the open boundary condition that is frequently symbolised as a hallmark of the skin-effect, is a system-size dependent phenomena, and appears even when the associated energies are indeed complex.
 	In addition, it is demonstrated that the spin-flip term in the RSO interaction infact possesses a tendency to diminish the directionality of the skin-effect. 
	On scrutinizing the dynamical attributes in our non-Hermitian system, we unveil that in spite of the fact that the spectral DL transition accords with the dynamical phase transition, interestingly, the system comes across hyper-diffusive and negative diffusion dynamical regimes depending upon the strength of the RSO interaction, in the spectrally localized regime.
	\end{abstract}
	
	\maketitle
	
	\section{Introduction}
	   The pioneering work of Anderson localization induced by random disorder in 1958 \cite{Anderson} has led its way
	   to fascinating discoveries in several domains of physics including photonics \cite{Sarychev,Lahini_2008,Jovic,Qiao}, acoustics \cite{Condat,Cohen}, superconductors \cite{Katsanos_1998,Burmistrov,Leavens}, solitonics \cite{Conti,Sacha_2009}, and a wide range of condensed matter systems \cite{Ching,Deutscher,Prelov}.
	   It was argued from the scaling theory of localization that the wave propagation through a random medium is completely suppressed even for an arbitrarily small disorder for systems below the critical dimension  ($d_c=3$), as a consequence of the interference between the scattering waves \cite{Abrahams}.
	   In contrast to the random lattices described by the Anderson Hamiltonian, the Aubry-Andr\'{e}-Harper(AAH) model \cite{Aubry,Harper,Sokoloff} stands as a paradigmatic example of quasicrystals which 
	   exhibit a delocalization-localization (DL) transition separating the metallic and the insulating states, even in one-dimension (1D).\\
	   \indent
	   The AAH Hamiltonian for closed quantum systems have received extensive attention from the researchers in several works in the last few years \cite{Tong,Ossipov,Yoo,Wang}.
	   On the other hand, a wide range of physical systems, for example, the cold-atoms \cite{Lee_2014,Li} are seldom isolated, and are typically described by an 
	   effective non-Hermitian Hamiltonian.
	   Since long, such model Hamiltonians have been used in exemplifying decaying states \cite{Faisal,Feshbach}.
	   In addition, in the last few years, the consequences of dissipation and drive in open quantum 
	   systems have been explored \cite{Sergi,Tripathi,Fukui,Oka}.
	   Such non-Hermitian systems have also been studied
	   in the context of many-body localization \cite{Matsumoto,Zhai,Lee}.
	   In the recent years, a great amount of investigations have been carried out on the localization, spectral and topological properties, self-duality, mobility edges,  and transport in
	   non-Hermitian quasiperiodic lattices \cite{Zeng,Liu,Zhai,Tzortzakakis,Tang}, while the investigation of 
	   real-time dynamics has been a principal objective in a wide range of non-Hermitian systems \cite{Baker,Lee,Kozlowski,WChen,Fehske}.\\
	   \indent
	   The interplay of non-Hermiticity and disorder can lead to several fascinating results, for example, the prototypical non-Hermitian Hatano-Nelson Hamiltonian 
	   with Anderson-type disorder and asymmetric hopping,  
	   results in a DL transition \cite{HatanoNelson1996,HatanoNelson1998,Kolesnikov} even in 1D, in contrast to the original Hermitian Anderson Hamiltonian.
	   Furthermore, it is established that in the non-Hermitian systems with uncorrelated disorder, the DL transition accompanies a simultaneous change in the energy spectrum from 
	   complex to real \cite{HatanoNelson1996,HatanoNelson1998}. In Refs.~\cite{HatanoTR} and \cite{Hatano_2021}, it is argued that the complex-conjugate pair of energy eigenvalues are manifestations of 
	   broken time-reversal symmetry of the eigenstates.
	   A similar DL transition is also possible when the random potential is replaced with a quasiperiodic on-site disorder. 
	   These non-Hermitian systems with asymmetric hopping offer uniqueness wherein the boundary conditions can 
	   drastically affect the bulk properties.
	   In particular, under the periodic boundary condition (PBC), all the eigenstates are extended 
	   below the critical point, whereas for an open boundary condition (OBC) these eigenstates localize at the edges of the system.
	   This feature, often dubbed as the non-Hermitian skin effect, violates the bulk-boundary correspondence observed in the 
	   Hermitian systems \cite{Lee_2016,Alvarez,Kunst,Yao,Evardsson}. 
	   Such a localization of macroscopic number of eigenstates at the boundary is often associated with the emergence of a real energy 
	   spectrum, and is considered as a signature of skin-effect. 
	   It is established that under the PBC, the energies form a loop in the complex plane, possessing a point-gap and a non-trivial topology \cite{Kawabata,Sato}.
	   However, under the open boundary condition, the spectrum becomes real, no longer retaining the point-gap, and the system becomes topologically trivial.
	   On the other hand, since long, several works have demonstrated the inter-dependence of the spectral properties to the dynamical propagation of the 
	   wave-packet in Hermitian systems\cite{Ostlund,Geisel,Ketzmerick,Ketzmerick1997,Deng}. The concurrent dynamical and spectral phase transitions have 
	   also been identified recently in non-Hermitian quasicrystals \cite{Longhi}. It is then natural to ask whether this correspondence
	   of the spectral and the dynamical properties to the DL transitions holds true under all circumstances.\\ 
	   \indent 
	   To address this question, in this work, we consider a time-reversal symmetric non-Hermitian counterpart of the tight-binding AAH 
	   Hamiltonian with Rashba Spin-Orbit (RSO) interaction. The time-reversal symmetry of the Hamiltonian remains unaffected with the addition of RSO interaction.
	   Surprisingly, as contrasted to the situation without RSO interaction, we have observed that in the presence of RSO, the eigenstates possess complex energies, even in the spectrally localized regime.
	   Moreover, we find evidence of the non-Hermitian skin-effect 
	   with complex energies under OBC, in contrary to the previous observation of the skin-effect that has frequently been associated to the reality of the spectrum.
	   Interestingly, we have observed that the connection of the skin-effect to the reality in the energy spectrum is a system-size dependent behavior in the systems without RSO interaction.
	   Furthermore, we find evidences that the directionality of the skin-effect significantly diminishes when the amplitude of the spin-flip is greater than the spin-conserving term in the RSO Hamiltonian.
	   In addition, we have observed that although the spectral and the dynamical phase transitions coincide in the Hamiltonian considered in this work, the interplay of the 
	   non-Hermiticity with the RSO interaction leads to exotic dynamical features in the spectrally localized regime. 
	   In particular, on estimating the diffusive exponent, we have observed dynamical delocalization behavior with hyper-diffusive
	   transport in the localized regime when the RSO interaction amplitudes in the non-Hermitian Hamiltonian are non-vanishing, 
	   in contrast to the Hermitian counterpart of the same Hamiltonian.
	   In addition, we have obtained a negative diffusion exponent in the spectrally localized regime in the presence of strong RSO interaction.
	   Finally, using the measure of Shannon entropy that has often been informative in the notion of steady state equilibrium \cite{Mandal, Narayanan},	
	   we have inferred that beyond the DL transition, there exists metastable state(s) prior to reaching the final equilibrium.
	   \\
	   \indent
	   To the best of our knowledge, such a complex-complex spectral transition concurrent with the DL transition,
	   and the existence of skin-modes possessing complex energies (under OBC) have not been reported in the non-Hermitian systems so far.
	   Moreover, the spin-dependent hyper-diffusive transport and the negative diffusion traits remain less explored till date.
	   The studies of the spin-dependent transport have played a central role 
	   in fabricating spintronic devices in heterostructures, chaotic quantum dots and semiconductors \cite{Egues,Beenakker,Datta}. 
	   In this line, our results in non-Hermitian systems will be beneficial for experimental realizations of such devices in spin-photonic
	   lattices \cite{Cardano, Zhang, Longhi2015}. \\
	   \indent
	   The rest of the work is organized as follows: In Sec.~\ref{Sec:Model}, we define our time-dependent non-Hermitian Hamiltonian in the presence of RSO interaction,
	   followed by the analysis of the energy spectrum in Sec.~\ref{Sec:IPR}.
	   We have discussed the skin-effect for our system in details in Sec.~\ref{Sec:Skin_effect}.
	   Sec.~\ref{Sec:Dynamical_study} elaborately demonstrates the dynamics of an excited spin-up fermion and its propagation in the lattice with time.
	   In Sec.~\ref{Sec:MSD_and_diffusivity}, we discuss the Mean Square Displacement (MSD) and the wave-packet spreading velocity, along with the estimated diffusion exponents for various 
	   strengths of the RSO interaction. Sec.~\ref{Sec:Shannon_entropy} is devoted to determine the stability using the Shannon entropy in the spectrally localized regime for different modulations of the RSO interaction.
	   We conclude our results and main findings of this work in Sec.~\ref{Sec:Conclusions}.
	   
		\begin{figure*}[]
		\centering 
		\includegraphics[width=0.24\textwidth,height=0.24\textwidth]{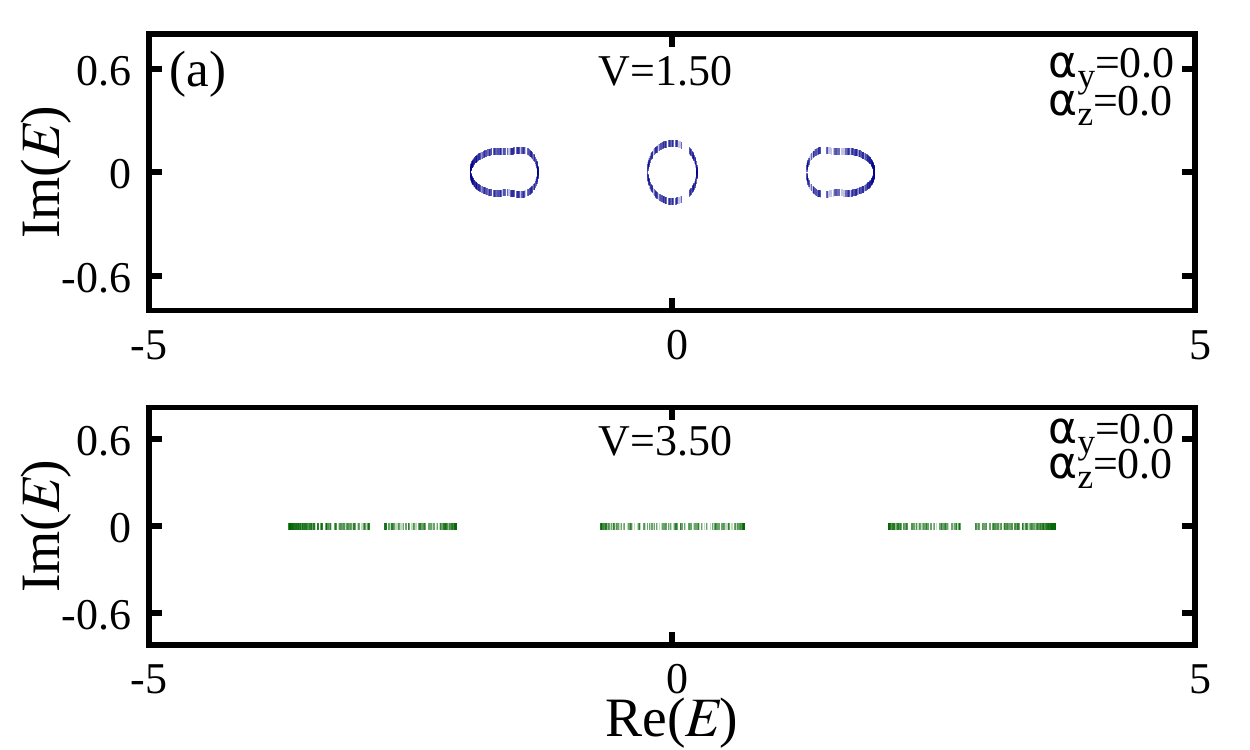}\hspace{0.0cm}
		\includegraphics[width=0.24\textwidth,height=0.24\textwidth]{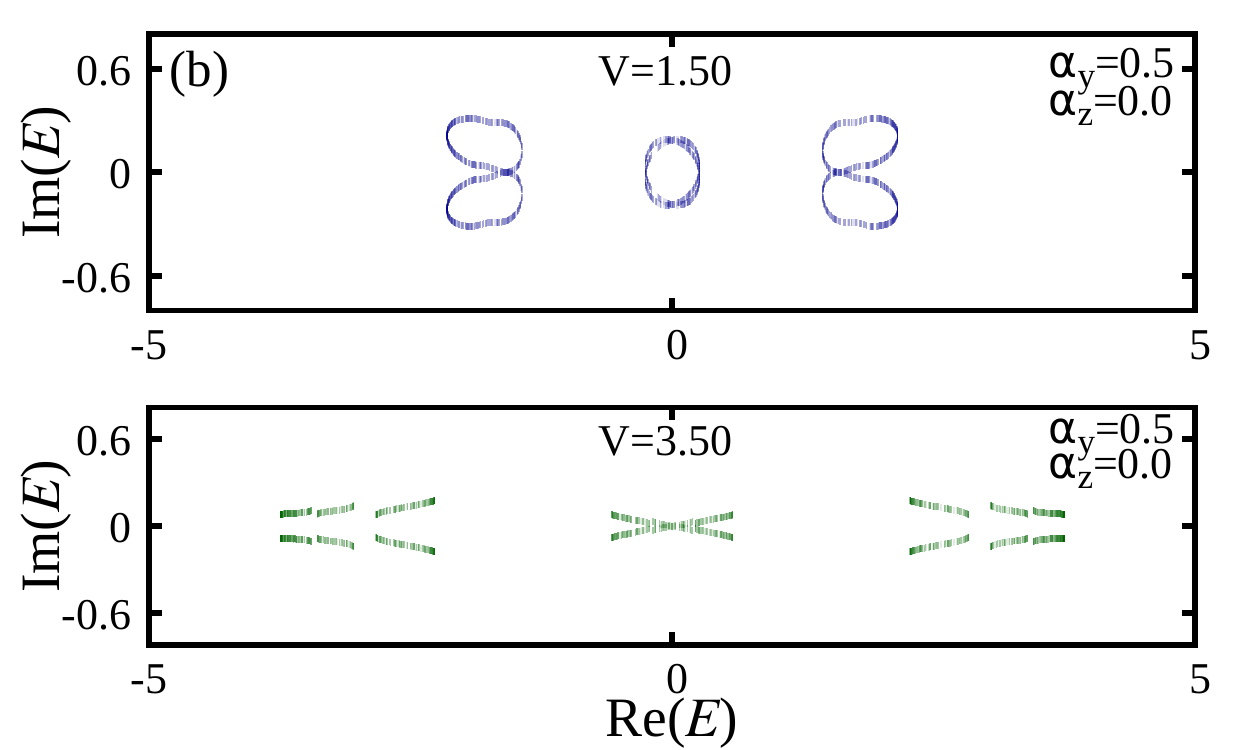}\hspace{0.0cm} 
		\includegraphics[width=0.24\textwidth,height=0.24\textwidth]{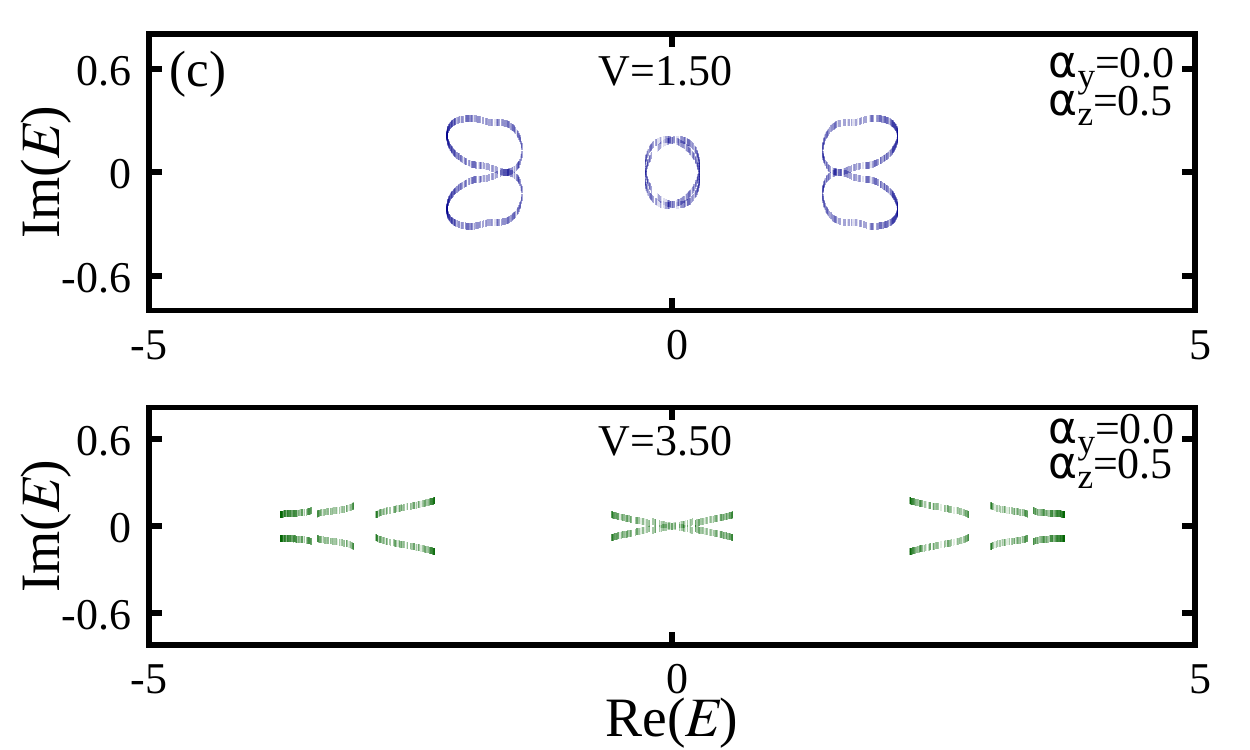}\hspace{0.0cm}
		\includegraphics[width=0.24\textwidth,height=0.24\textwidth]{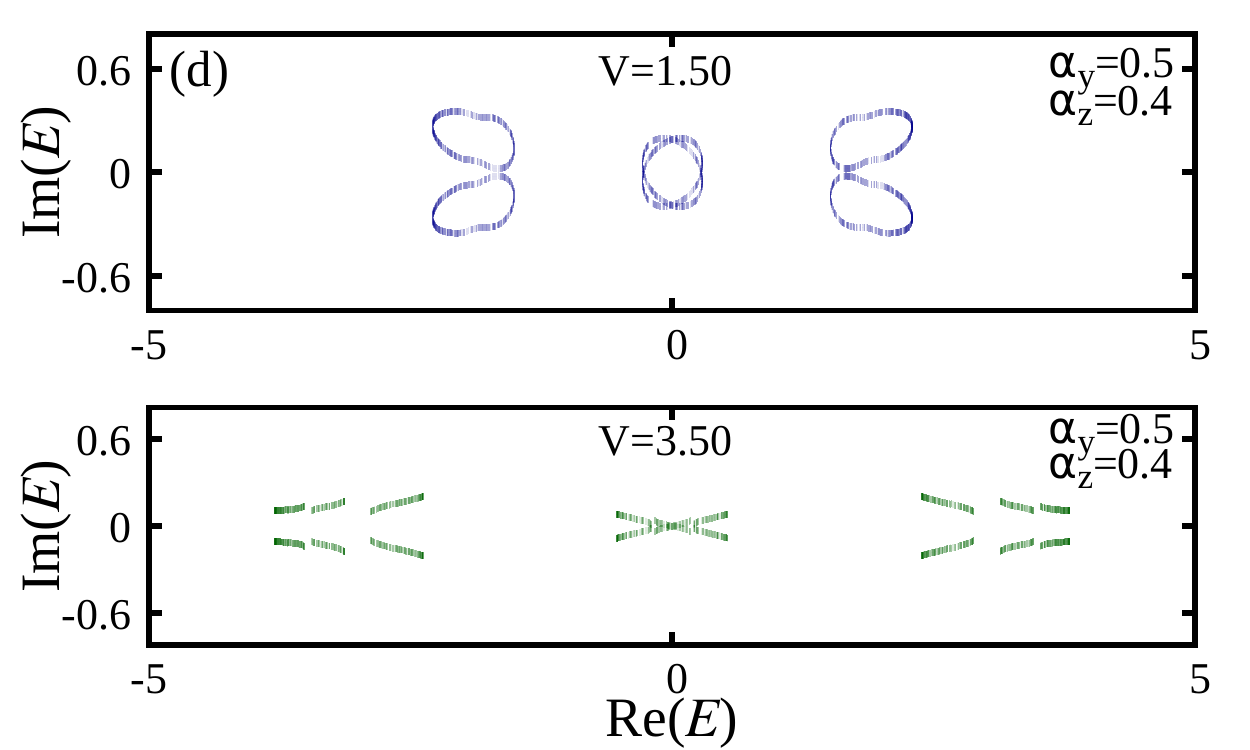}
		\vspace{-0.2cm}
		\caption{ The energy spectrum in the complex plane for the four distinct non-Hermitian $(J_L/J_R=0.5)$ cases: 
		(a) without the RSO interaction ($\alpha_y=0.0$ and $\alpha_z=0.0$),
		(b) with the spin-conserving RSO interaction ($\alpha_y=0.5$ and $\alpha_z=0.0$),
		(c) in the presence of the spin-flip part of the RSO interaction ($\alpha_y=0.0$ and $\alpha_z=0.5$), and
		(d) in the presence of both the RSO interaction terms ($\alpha_y=0.5$ and $\alpha_z=0.4$).
		The dark-blue and dark-green markers depict the energy spectrum in the spectrally delocalized and localized regimes respectively, as determined in Appendix~\ref{App:IPR}.
		Here, we have employed the periodic boundary condition on a lattice with 610 sites.}
		\label{Fig:Energy_spectrum}
	\end{figure*}
	
	\section{The non-Hermitian quasi-periodic Hamiltonian with Rashba spin-orbit interaction}
	
	\subsection{The non-Hermitian model}\label{Sec:Model}
	The Hamiltonian considered in this work is a non-Hermitian extension of the one-dimensional AAH model \cite{Aubry, Harper}
	with a quasiperiodic potential that is incommensurate to the underlying lattice period, and in the presence of the RSO interaction. 
	Similar to the Hatano-Nelson Hamiltonian, the non-Hermiticity arises due to the asymmetric hopping amplitudes.
	Recently, the dynamical phase transitions in such a protypical model without the RSO interaction has been discussed in Ref.~\cite{Longhi}.
	In the presence of RSO interaction \cite{Birkholz, Mireles}, the Hamiltonian is defined as,
	\begin{eqnarray}
	\mathcal{H}= J_R\displaystyle\sum_{n,\sigma} c^\dag_{{n+1},{\sigma}} c_{{n},{\sigma}}
	+J_L\displaystyle\sum_{n,\sigma} c^\dag_{{n},{\sigma}} c_{{n+1},{\sigma}} ~~~~~~~~~~~\nonumber \\
	+\sum_{n,{\sigma}} V_n c^\dag_{{n},{\sigma}} c_{{n},{\sigma}}
	-\alpha_z \displaystyle\sum_{n,{\sigma},{\sigma'}}(c^\dag_{n+1,\sigma}(i\sigma_y)_{\sigma,\sigma'}
	c_{n,\sigma'}+ h.c.) \nonumber \\
	+\alpha_y \displaystyle\sum_{n,{\sigma},{\sigma'}} (c^\dag_{n+1,\sigma}(i\sigma_z)_{\sigma,\sigma'}  
	c_{n,\sigma'}+ h.c.),~~~~~~~~~~~~~~~~~~~~~~~~\label{Eq:Total_Hamiltonian}
	\end{eqnarray}
 	where $J_R$ and $J_L$ indicate the nearest neighbor hopping amplitudes when the electrons move
 	towards the right and left directions respectively.
 	$N$ is the number of sites in the lattice, and $n$ denotes the lattice site index.
 	$N$ is approximated as $N=F_{n-1}/F_n$, where $F_{n}$ and $F_{n-1}$ are the $n^{th}$ and $(n-1)^{th}$ numbers in the Fibonacci sequence.
 	$c^\dag_{n,\sigma}$ and $c_{n,\sigma}$ are the fermionic creation and annihilation operators with spin-configuration $\sigma(\uparrow,\downarrow)$.
 	$\alpha_y$ and $\alpha_z$ denote the spin-conserving and spin-flip hopping amplitudes due to the RSO interaction,
 	whereas $\sigma_y$ and $\sigma_z$ are the components of Pauli's spin-matrices.

 	The quasi-periodic on-site potential $V_n$ in Eq.~(\ref{Eq:Total_Hamiltonian}) is given by,
 	\begin{eqnarray}
 		V_n=V \text{cos}(2\pi\alpha n),  \nonumber
 	\end{eqnarray}
 	where $\alpha$ is the inverse of the golden ratio given by $\alpha=(\sqrt5-1)/2$.
 	After setting $\ket{\psi(t)}=\sum_{n,\sigma} a_n^{\sigma}(t)\ket{n^{\sigma}}$, the time-dependent coupled Schr\"{o}dinger equations for the amplitudes ($a_n$) 
 	of the two spin orientations derived from Eq.~(\ref{Eq:Total_Hamiltonian}) can be written as,
	\begin{eqnarray}
	\mathcal{H}a_{n}^{\uparrow}=(J_R+i\alpha_y)a_{n-1}^{\uparrow} +(J_L-i\alpha_y)a_{n+1}^{\uparrow}~~~~~~~~~~~~~~~~~~~& &\nonumber \\
	+\alpha_z \left(a_{n+1}^{\downarrow} - a_{n-1}^{\downarrow} \right)
	+ V~{\rm{cos}}(2\pi\alpha \textit{n}) a_{n}^{\uparrow} = i \frac{\emph{d}a_n^{\uparrow}}{\emph{dt}} , ~~~~~~~~~~~~\label{Eq:Spin_up_eqn}\\
	\text{and}~~~~~~~~~~~~~~~~~~~~~~~~~~~~~~~~~~~~~~~~~~~~~~~~~~~~~~~~~~~~~~~~~~~~~ \nonumber \\
	\mathcal{H}a_{n}^{\downarrow}=(J_R-i\alpha_y)a_{n-1}^{\downarrow} +(J_L+i\alpha_y)a_{n+1}^{\downarrow}
	~~~~~~~~~~~~~~~~~~~& &\nonumber \\
	-\alpha_z \left(a_{n+1}^{\uparrow} - a_{n-1}^{\uparrow} \right)
	+ V~{\rm{cos}}(2\pi\alpha \textit{n}) a_{n}^{\downarrow} = i \frac{\emph{d}a_n^{\downarrow}}{\emph{dt}} .~~~~~~~~~~~~\label{Eq:Spin_down_eqn}
	\end{eqnarray}
	
	\begin{figure*}[]
		\centering 
		\includegraphics[width=0.24\textwidth,height=0.24\textwidth]{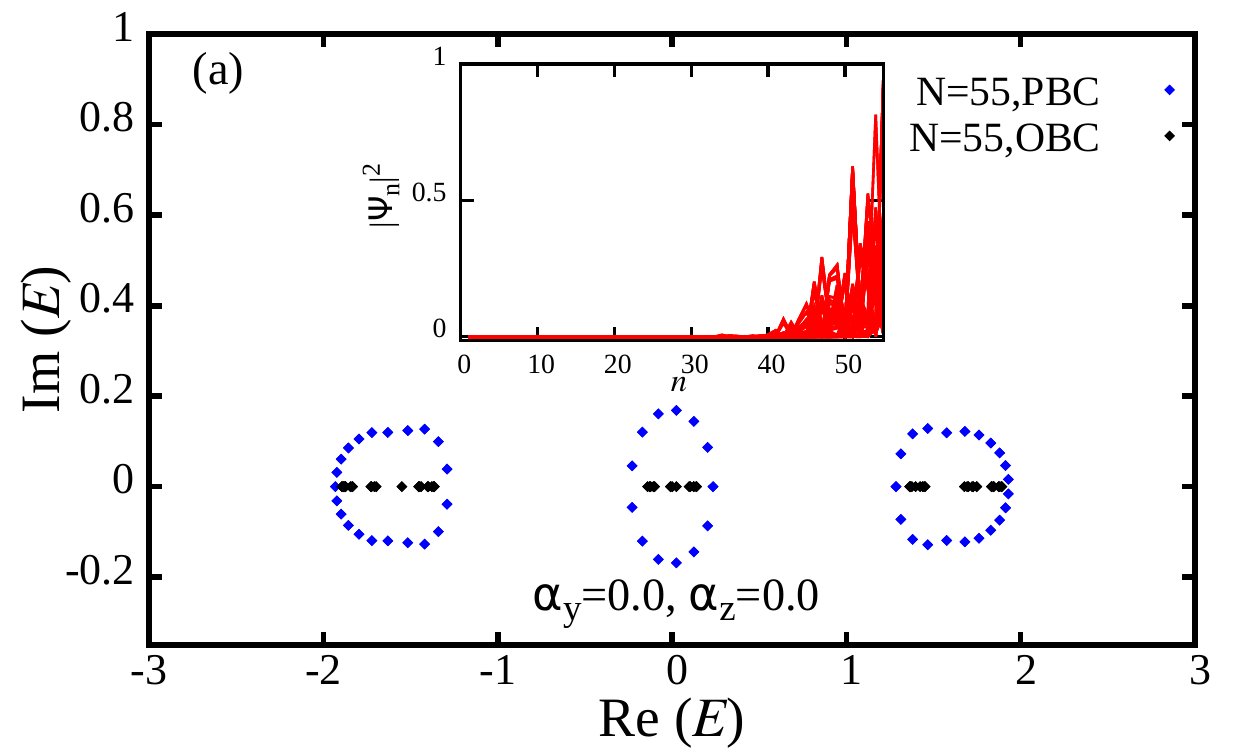}\hspace{0.0cm}
		\includegraphics[width=0.24\textwidth,height=0.24\textwidth]{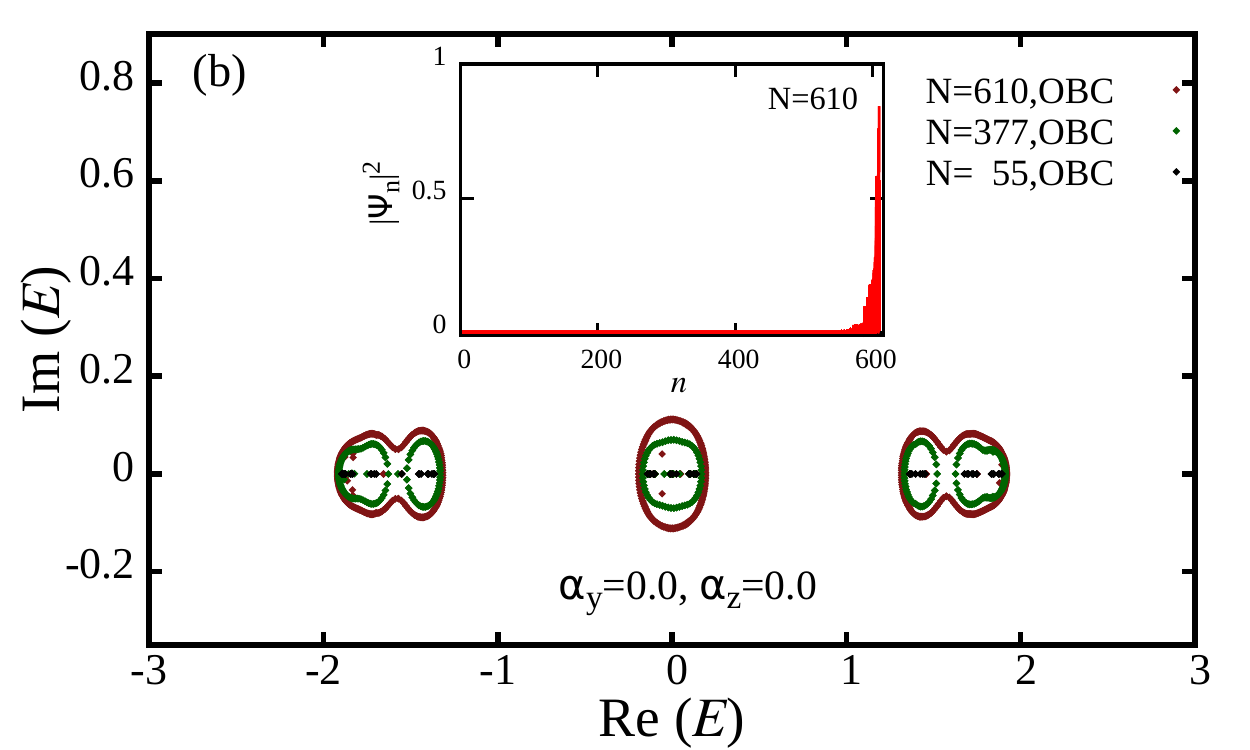}\hspace{0.0cm} 
		\includegraphics[width=0.24\textwidth,height=0.24\textwidth]{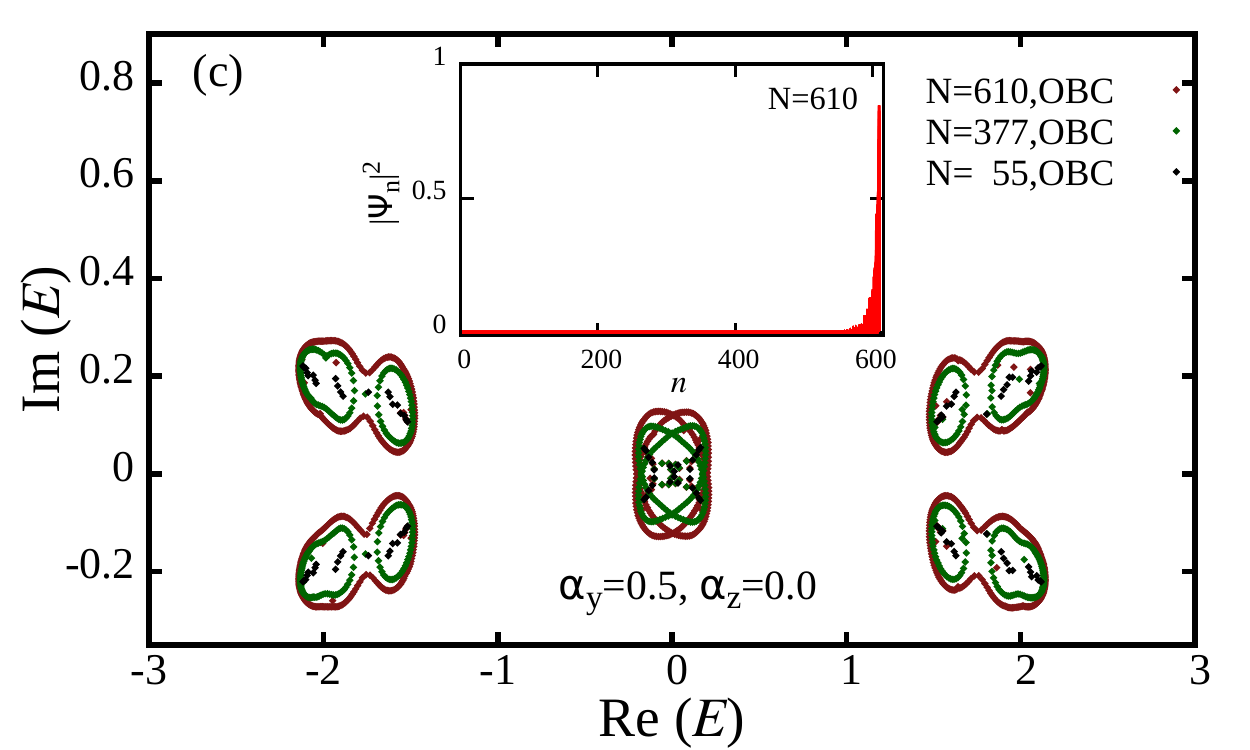}\hspace{0.0cm}
		\includegraphics[width=0.24\textwidth,height=0.24\textwidth]{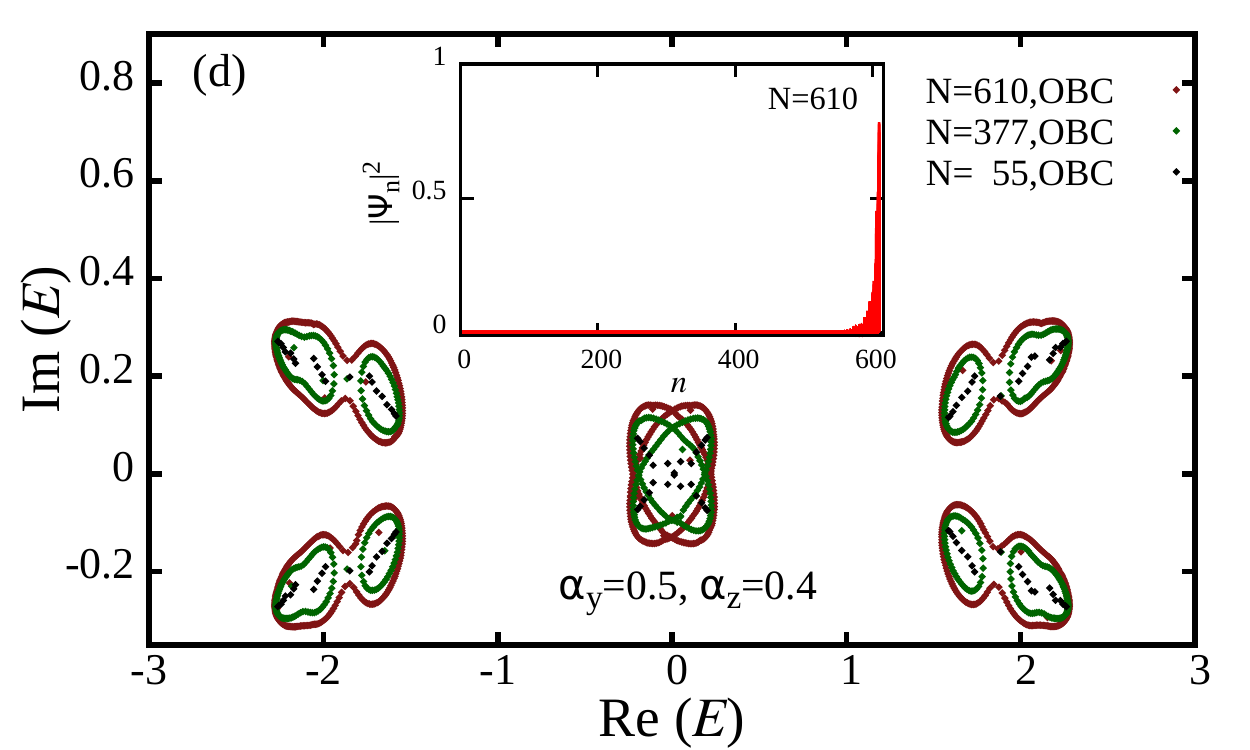}
		\vspace{-0.2cm}
		\caption{ The energy spectrum of the Hamiltonian given in Eq.~(\ref{Eq:Total_Hamiltonian}) with $J_L/J_R=0.5$, along with the density profile 
		of the eigenstates (insets) in the delocalized regime ($V=1.50$).
		(a) The complex energy spectrum for a non-reciprocal lattice without RSO interaction ($\alpha_y=0.0$ and $\alpha_z=0.0$) with 55 sites under PBC (in blue),
		and the corresponding real energy spetrum under OBC (in black), according to Refs.~\cite{Lee_2019,Sato}. 
		The inset shows the skin modes localized at the right end.
		(b) The system size dependence of the energy spectrum (under OBC) for the same set of parameters as in (a).
		The inset shows the existence of skin modes, corresponding to 610 sites.
		(c) The energy spectrum (under OBC) and the existence of skin modes in the presence of only the spin-conserving RSO interaction term 
		($\alpha_y=0.5$ and $\alpha_z=0.0$).
		(d) The energy spectrum (under OBC) and the existence of skin modes in the presence of both the spin-conserving and spin-flip RSO interaction 
		terms ($\alpha_y=0.5$ and $\alpha_z=0.4$).
 		In the last three panels, the energy spectrum is plotted for $N=55,377$ and $610$, which are shown in black, green and brown respectively.}
		\label{Fig:Skin_effect}
	\end{figure*}

	\subsection{Delocalization-localization transition and the energy spectrum}\label{Sec:IPR}
	The special case of Hermiticity sets in when $J_L/J_R=1$ in Eq.~(\ref{Eq:Total_Hamiltonian}). In such systems,
	the DL transition occurs at the critical point $V_c=2J_{L/R}$
	\cite{Aubry}, and the energy spectrum remains real in both the delocalized and localized regimes with RSO interaction. 
	We introduce the non-Hermiticity by considering $J_L/J_R=0.5$. All the parameters of the Hamiltonian is measured
	with respect to $J_R$, where we set $J_R=1$. 
	In Ref.~\cite{Longhi}, it is demonstrated that for such systems (for any value of $J_L$, provided that $J_R>J_L$), the DL phase
	transition occurs at the critical value of the quasiperiodic potential given by,
	\begin{eqnarray}
		V_c=2J_R.
	\end{eqnarray}
	
	In the presence of RSO interaction, we expect a similar transition at a rescaled value of $J_R$(depending upon the strength of the RSO interaction). 
	In Appendix~\ref{App:IPR}, we have obtained the critical value $V_c$ of the spectral transition both with and without RSO interactions,
	which will be used for the discussions hereafter.\\
	\indent
	The Hamiltonian in Eq.~(\ref{Eq:Total_Hamiltonian}) is symmetric under the time-reversal, given by the condition  $\mathcal{TH^*T}^{-1}=\mathcal{H}$ \cite{Kawabata}, 
	where $\mathcal{T}$ is the time-reversal symmetry operator defined as $\mathcal{T}=i\sigma_y\otimes\mathbb{I}$.
	In time-reversal symmetric non-Hermitian Hamiltonians, the existence of the symmetry in the eigenstates is associated with the reality of the corresponding
	energy eigenvalues, whereas the energies for the symmetry-breaking eigenstates occur in complex conjugate pairs \cite{HatanoTR,Hatano_2021,Sudarshan}.
	In both the Hermitian and the non-Hermitian quasiperiodic Hamiltonians, the exponentially localized eigenstates are illustrated to have a real point 
	spectrum \cite{Jitomirskaya,Avila,Longhi}.
	Fig.~(\ref{Fig:Energy_spectrum}) portrays the energy spectrum in the complex plane for the Hamiltonian defined in Eq.~(\ref{Eq:Total_Hamiltonian}), for different strengths of the RSO interaction.
	In the absence of the RSO interaction (Fig.~\ref{Fig:Energy_spectrum}(a)), it is established that under the periodic boundary conditions,
	below the critical point $V_c= 2.0$, the states remain extended and the energy spectrum is complex (eigenstates break the time-reversal symmetry), 
	similar to the Hatano-Nelson model with random potentials \cite{HatanoNelson1996,HatanoNelson1998}. 
	The spectrum becomes real (eigenstates preserve the time-reversal symmetry) after the DL transition. In the presence of a single RSO interaction hopping amplitude
	($\alpha_y = 0.5, \alpha_z =0$ and vice-versa) the DL transition is pushed to a higher value of $V_c \simeq 2.24$. 
	However, in contrast to the system without RSO interaction, in this case,
	the energy spectrum remains complex even after the DL transition (Figs.~\ref{Fig:Energy_spectrum}(b-c)).
	It is evident that although the non-Hermitian Hamiltonian with RSO as defined in Eq.~(\ref{Eq:Total_Hamiltonian}),
	retains its intrinsic symmetry under the time-reversal,
	the states spontaneously break the symmetry. 
	Moreover, it is clear from Figs.~\ref{Fig:Energy_spectrum}(b-c) that both the spin-conserving and the spin-flip interaction exhibit identical spectral attributes.
	Fig.~\ref{Fig:Energy_spectrum}(d) illustrates the spectrum in the presence of both the RSO interaction terms. In this case, the overall spectral features
	still remain the same. However, the critical point is pushed to an even higher value of $V_c \simeq 2.37$. 
	From Figs.~\ref{Fig:Energy_spectrum}(b-d), it is evident that in the presence of RSO interaction (even a single non-vanishing RSO coupling amplitude),
	the eigenstates remain in the broken time-reversal symmetric regime for all strengths of the quasiperiodic potential.
	Such a complex spectrum beyond $V_c$ is expected to lead to unusual dynamical features in the spectrally localized regime in systems with RSO interactions.
	
	\subsection{Skin effect in the non-Hermitian model with RSO}\label{Sec:Skin_effect}

	The skin-effect discovered in the recent years \cite{Lee_2016,Alvarez,Kunst,Yao,Evardsson}, reveals extreme sensitivity
	of the spectrum to the boundary condition in non-Hermitian systems.
	The bulk Bloch modes observed under the periodic boundaries localize at one of the edges of the system under the OBC, violating the bulk-boundary correspondence
	observed in Hermitian counterparts.
	It has been recently demonstrated \cite{Sato} that under the PBC, and without any disorder,
	the complex energy spectrum lies on an ellipse following a Fourier transformation (shown in Fig.~\ref{Fig:NH_skin_effect}(a) of Appendix~\ref{App:Skin_effect}).
	In contrast, under the OBC, the system can be mapped to a Hermitian analog via. an imaginary Gauge transformation and a similarity transformation,
	yielding completely real eigenspectrum whose eigenstates are localized at a boundary.
	Fig.~\ref{Fig:Skin_effect}(a) illustrates the behavior of the eigen spectrum under the PBC (in blue) and the OBC (in black), showing the transition of the spectrum 
	for a lattice with 55 sites in the presence of a disorder, and without the RSO interaction.  
	The eigenstates (called the skin modes) are localized at the right end as a consequence of the uni-directionality of the electrons towards the right.
	However, surprisingly, we find that the reality exists only for small system sizes (Fig.~\ref{Fig:Skin_effect}(b)),
	and the spectrum under OBC becomes complex as the number of sites in the lattice are increased.
	Furthermore, it is numerically verified that there exists skin modes even when the spectrum remains complex (shown in the inset of Fig.~\ref{Fig:Skin_effect}(b)), 
	in contrary to the observations of skin-effect due to the reality of eigenvalues as discussed in the previous works.
	Interestingly, in stark contrast to the situation without RSO interaction (Fig.~\ref{Fig:Skin_effect}(b)),
	in the presence of the RSO interaction we find that the eigenenergy spectrum is always complex under OBC, irrespective of the system size, as evident
	from Figs.~\ref{Fig:Skin_effect}(c-d). It is clear that the skin-effect persists even in the presence of the RSO interaction 
	(for different combinations of $\alpha_y$ and $\alpha_z$). Furthermore, we have verified that the observations of the skin-effect with complex eigenspectrum
	persist in the presence of RSO interaction, even in systems without disorder. The skin effect has been further discussed in Appendix~\ref{App:Skin_effect}.
	
	To obtain a clearer quantitative picture of the boundary localization, we employ a recently developed method of the directional-IPR (dIPR) defined as \cite{Zeng_2022},
	\begin{eqnarray}
		d\text{IPR}(\psi_j)=\mathcal{P}(\psi_j)\frac{\displaystyle\sum_{n,\sigma} |\psi_{n,j}^{\sigma}|^4}{(\displaystyle\sum_{n,\sigma} |\psi_{n,j}^{\sigma}|^2)^2},
		\label{Eq: dIPR}
	\end{eqnarray}
	Here, $\mathcal{P}(\psi_j)$ determines whether the states are localized towards the left or the right boundary depending upon the sign of the $sgn$ function, and is given as,
	\begin{eqnarray}
		\mathcal{P}(\psi_j)=\text{sgn}\Bigg[\displaystyle\sum_{n,\sigma}\Big(n-N/2-\delta'\Big)|\psi_{n,j}^{\sigma}|\Bigg].
		\label{Eq: dIPR_P}
	\end{eqnarray}
	
	$\delta'$ can take values in between 0 and 0.5.
	It is important to note that the dIPR is 0 when the states are delocalized, whereas for right (left) boundary localization, dIPR is positive (negative).
	The average of dIPR is then given as,
	\begin{eqnarray}
		d\text{MIPR}=\frac{1}{N} \displaystyle\sum_{j} \text{dIPR}(\psi_j).
		\label{Eq: dMIPR}
	\end{eqnarray}
	
\begin{figure}
		\vspace{-1.1cm}
		\begin{center}	
			\begin{tabular}{p{\linewidth}c}
				\begin{center}
					
					\includegraphics[width=0.34\textwidth,height=0.32\textwidth]{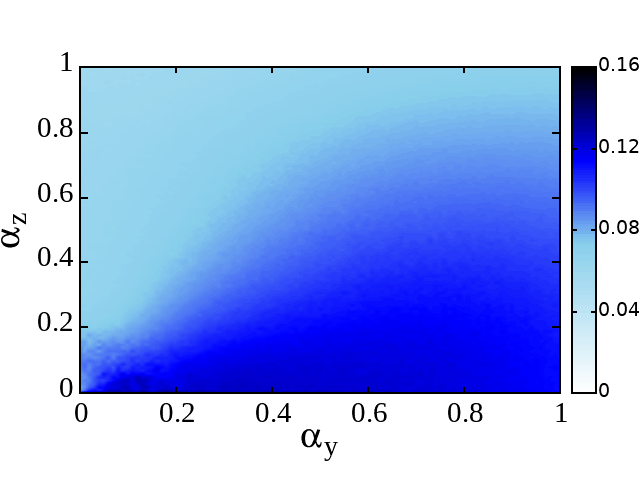}\hspace{0.0cm}
					
				\end{center}
			\end{tabular}	
		\end{center}
		\vspace{-1.2cm}
		\caption{Phase diagram of the dMIPR demonstrating the degree of boundary-localization due to  
		the skin-effect in the spectrally delocalized regime as a function of $\alpha_y$ and $\alpha_z$. We have used open 
		boundary condition on a lattice with 610 sites. Here $J_L/J_R=0.5$, and $V=1.50$.}
		\label{Fig:DIPR}
	\end{figure}

	The dMIPR ranges from +1 to -1, when either $J_L$ or $J_R$ vanishes, implying complete directionality in the system.
	We have presented the phase diagram of the dMIPR for different strengths of the RSO interaction ($\alpha_y$ and $\alpha_z$) in Fig.~(\ref{Fig:DIPR}).
	It is clear that when the system encounters a spin-flip interaction ($\alpha_z$) stronger than the spin-conserving interaction ($\alpha_y$), the directionality of 
	the skin-effect (towards the right boundary) becomes subdued, which is manifested by the light-blue section in the phase diagram.
		
	\begin{figure*}[]
		\centering 
		\includegraphics[width=0.28\textwidth,height=0.25\textwidth]{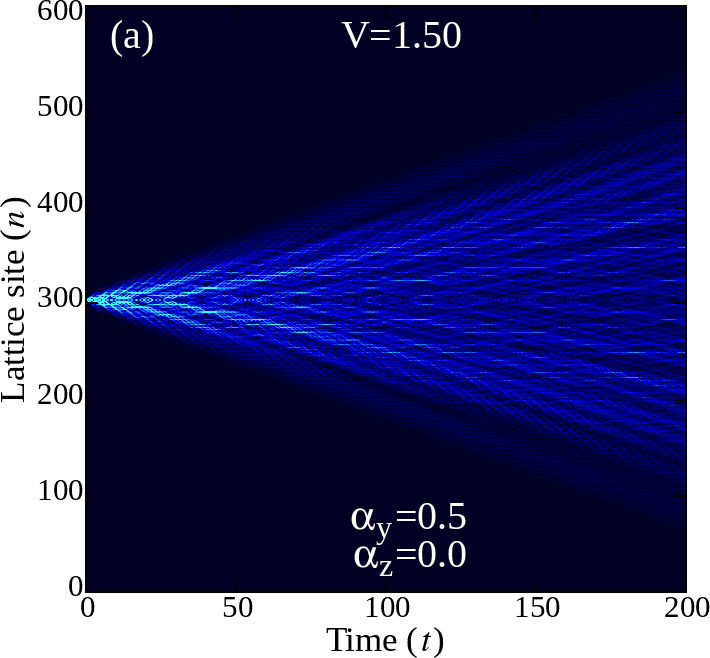}
		\includegraphics[width=0.28\textwidth,height=0.25\textwidth]{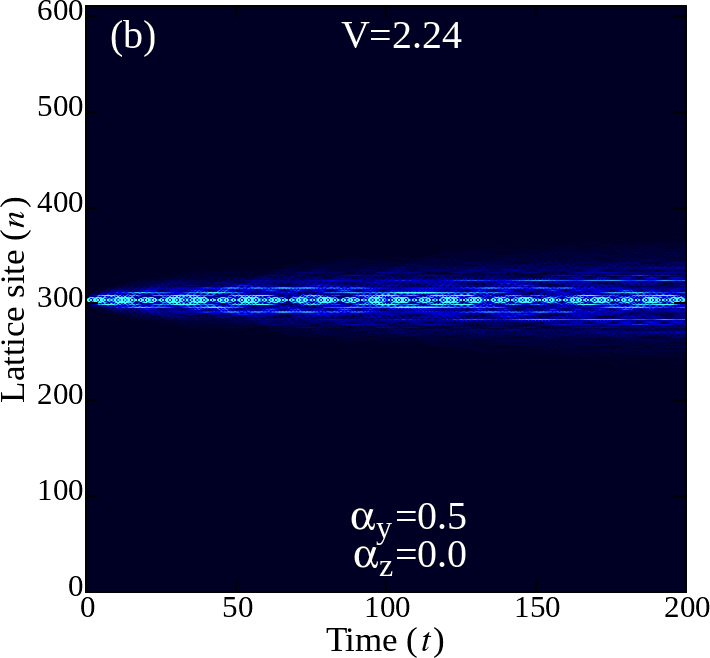}
		\includegraphics[width=0.28\textwidth,height=0.25\textwidth]{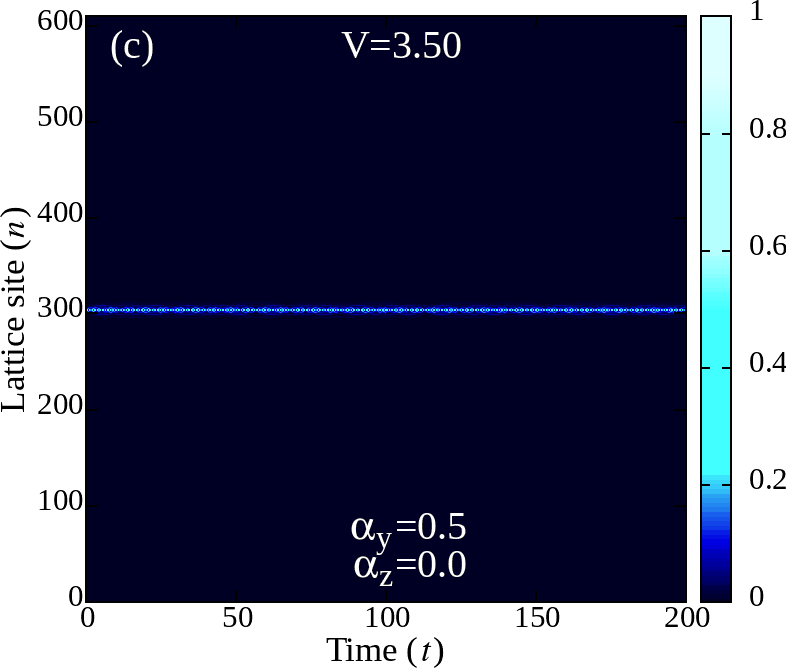}\\
		\includegraphics[width=0.28\textwidth,height=0.25\textwidth]{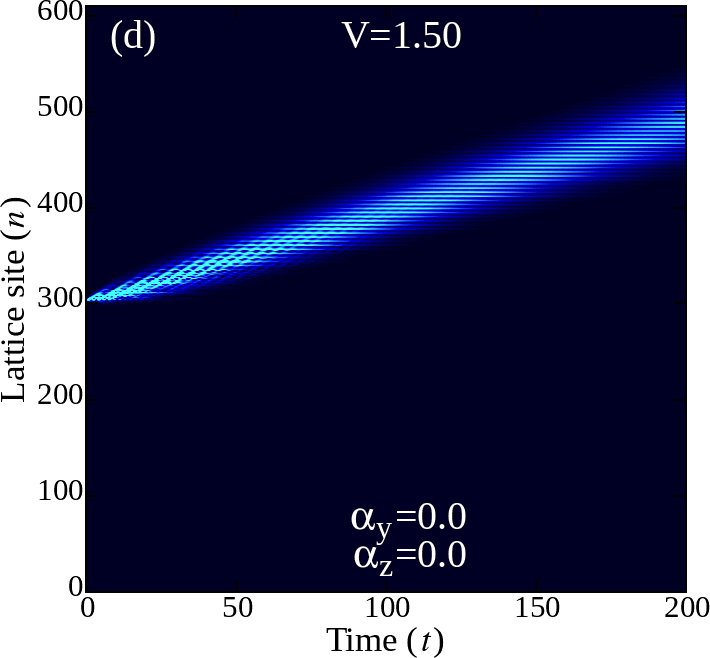}
		\includegraphics[width=0.28\textwidth,height=0.25\textwidth]{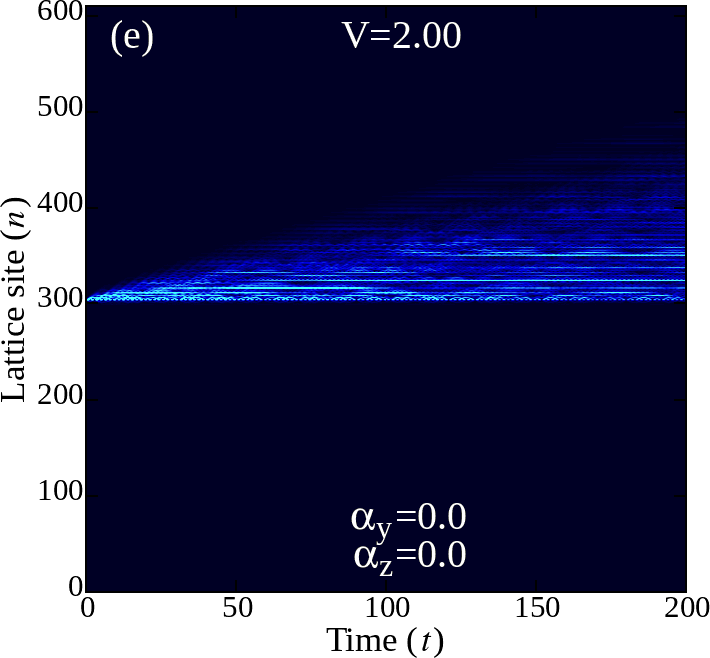}
		\includegraphics[width=0.28\textwidth,height=0.25\textwidth]{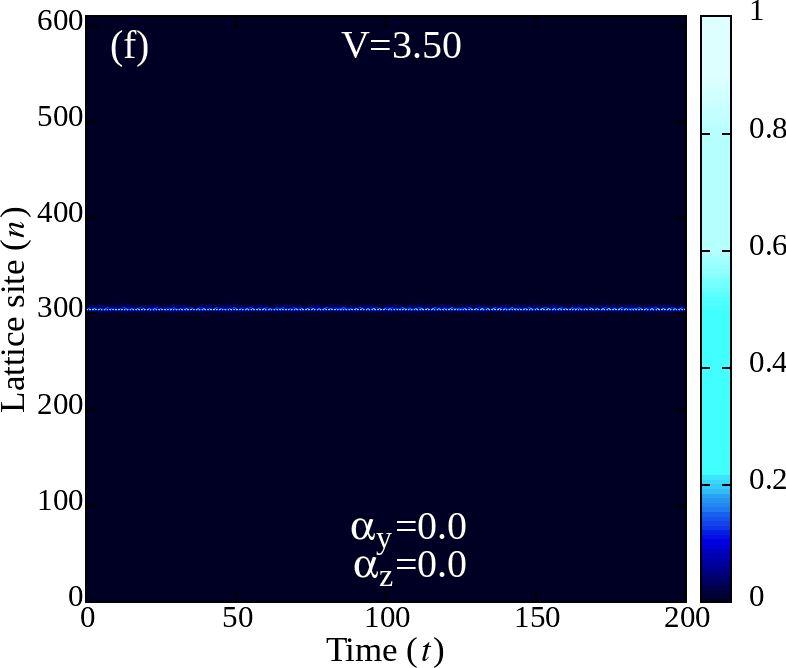}\\
		\vspace{0.2cm}
		\includegraphics[width=0.28\textwidth,height=0.25\textwidth]{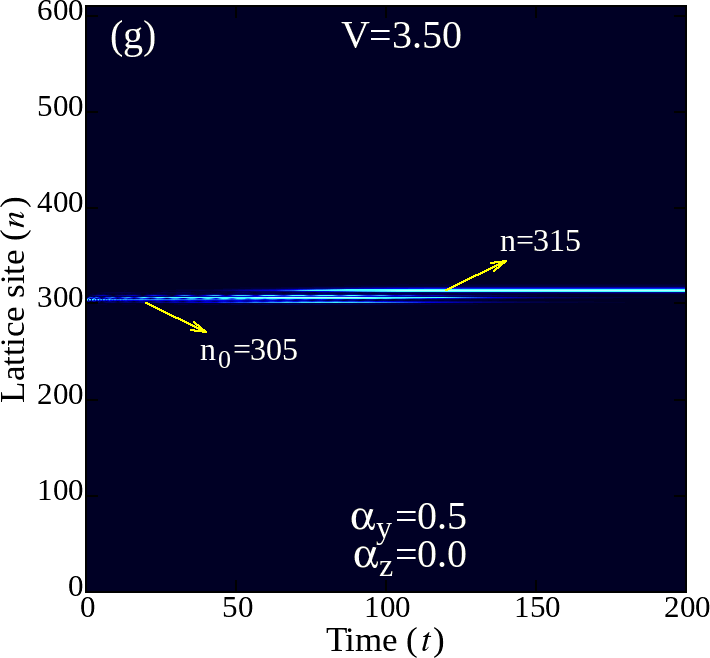}
		\includegraphics[width=0.28\textwidth,height=0.25\textwidth]{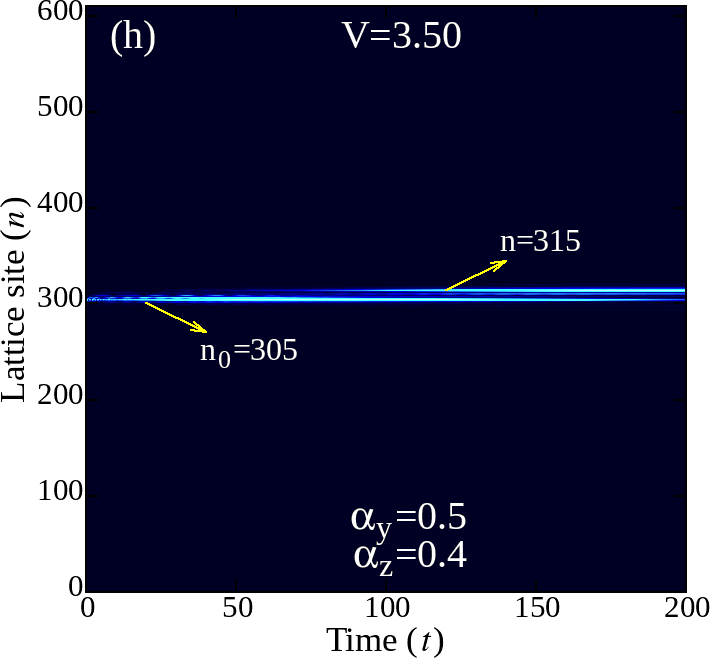}
		\includegraphics[width=0.28\textwidth,height=0.25\textwidth]{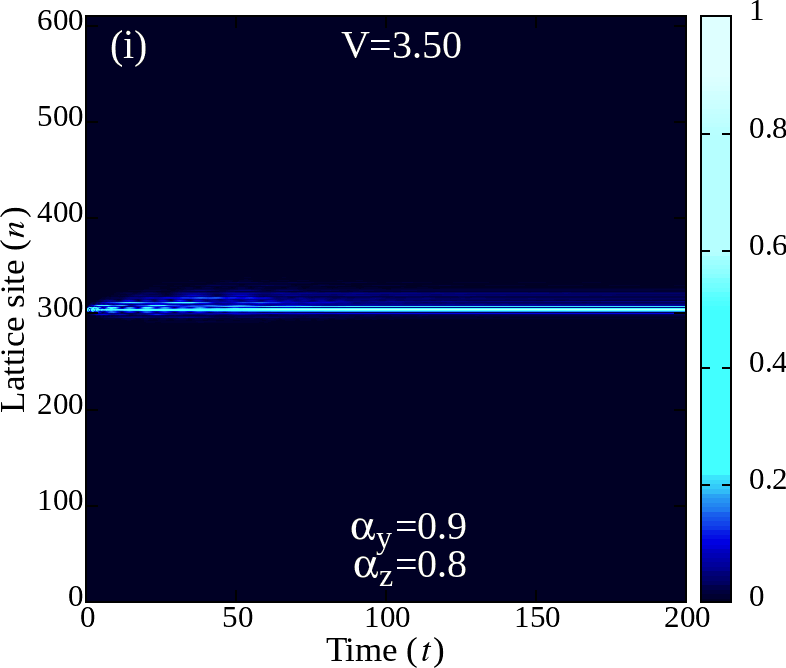}
		\caption{ The amplitude of the wave profile ($\vert\psi_n(t)\vert$) for different cases. (a-c) Hermitian system ($J_L=J_R=1$) with RSO interaction 
		($\alpha_y=0.5$ and $\alpha_z=0.0$). From left to right, the stengths of the quasiperiodic potential are V=1.50 (spectrally delocalized regime), 2.24 
		(critical point $V_c$ as determined in Appendix~\ref{App:IPR}) and 3.50 (spectrally localized regime) respectively.
		(d-f) Non-Hermitian system ($J_L/J_R=0.5$) without RSO interaction ($\alpha_y=0.0$ and $\alpha_z=0.0$). From left to right, the stengths of the quasiperiodic potential 
		are V=1.50 (spectrally delocalized regime), 2.00 (critical point) and 3.50 (spectrally localized regime) respectively. 
		(g) Non-Hermitian system ($J_L/J_R=0.5$) with only the spin-conserving RSO interaction term ($\alpha_y=0.5$ and $\alpha_z=0.0$) in the spectrally 
		localized regime (where $V_c \simeq 2.24$). 
		(h) Non-Hermitian system ($J_L/J_R=0.5$) with both the spin-conserving and the spin-flip RSO interaction terms ($\alpha_y=0.5$ and $\alpha_z=0.4$) in the 
		spectrally localized regime (where $V_c \simeq 2.37$). 
		(i) Non-Hermitian system ($J_L/J_R=0.5$) with strong RSO interaction ($\alpha_y=0.9$ and $\alpha_z=0.8$) in the spectrally localized regime (where $V_c \simeq 3.13$).
		In all the cases, we have considered a lattice of 610 sites and employed the periodic boundary conditions. Here, the wave-packet dynamics is demonstrated for 200 secs.}
		\label{Fig:Wave_profile}
	\end{figure*}
	
	\begin{figure}
		\begin{center}	
			\begin{tabular}{p{\linewidth}c}	
				
				\includegraphics[width=0.23\textwidth,height=0.22\textwidth]{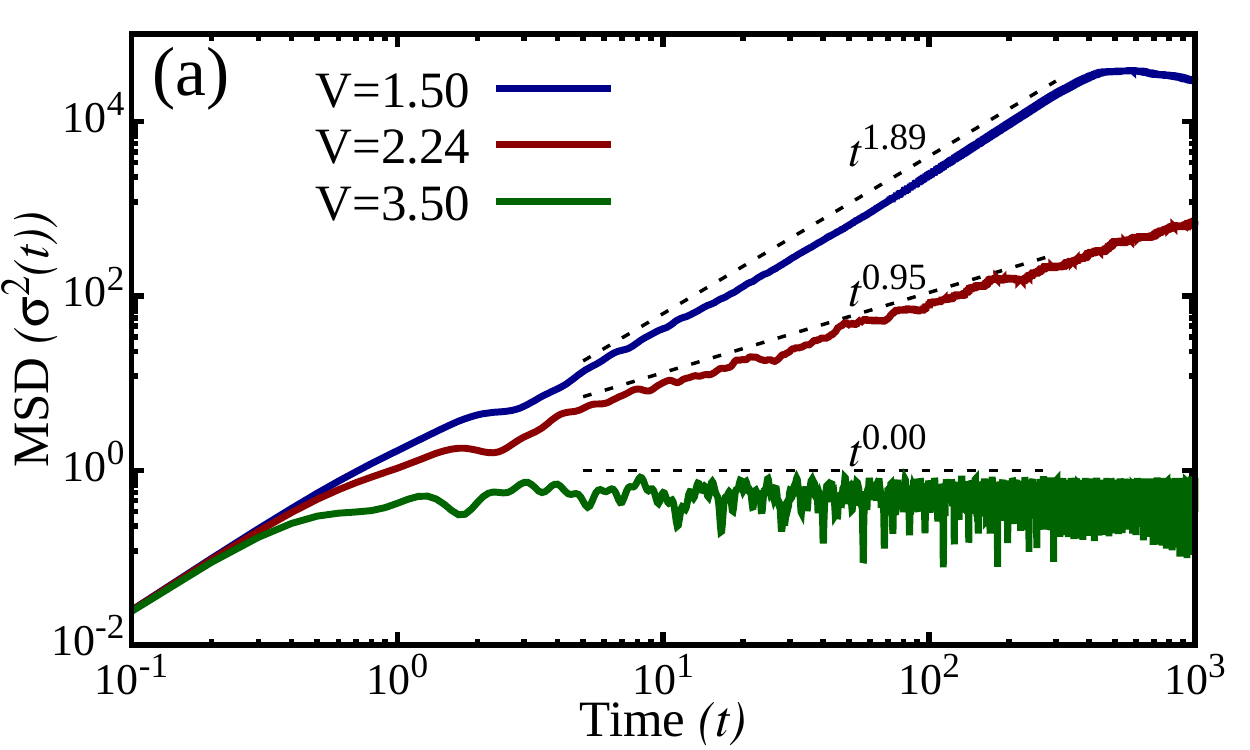} 	 \includegraphics[width=0.23\textwidth,height=0.22\textwidth]{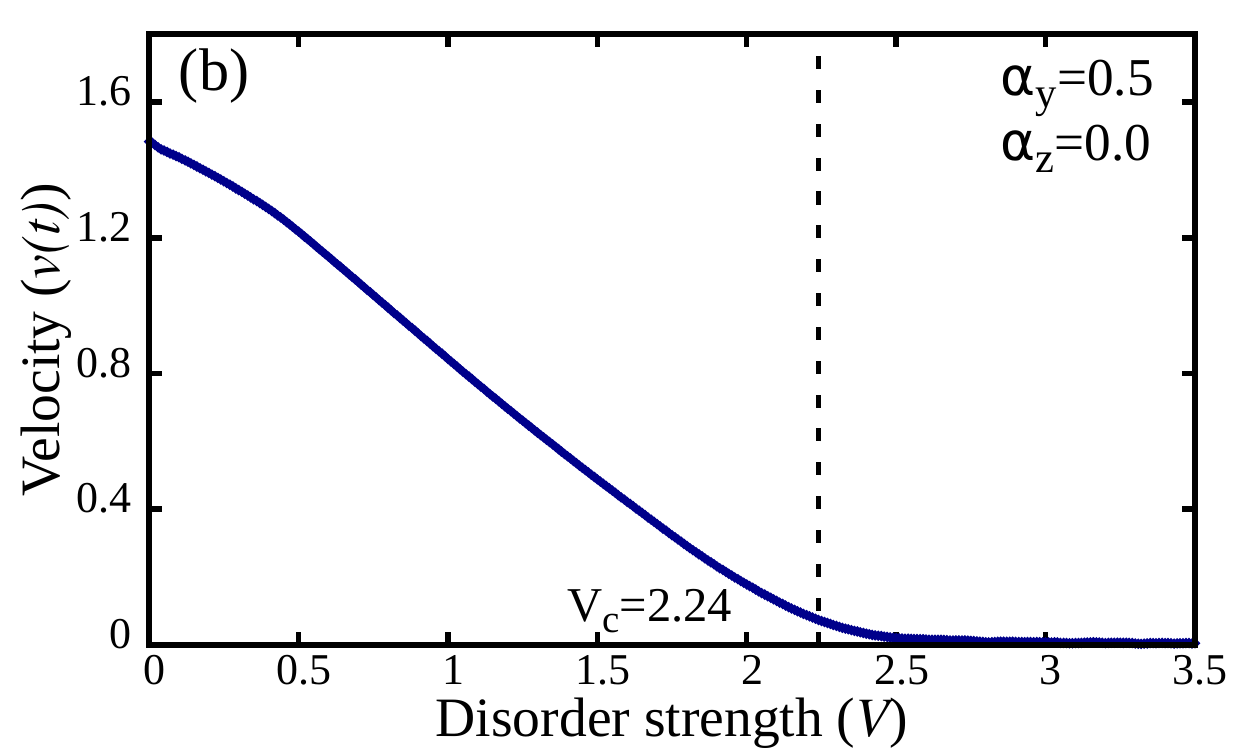}	
				
			\end{tabular}
			
		\end{center}
		\vspace{-1cm}
		\begin{center}	
			\begin{tabular}{p{\linewidth}c}	
				
				\includegraphics[width=0.23\textwidth,height=0.22\textwidth]{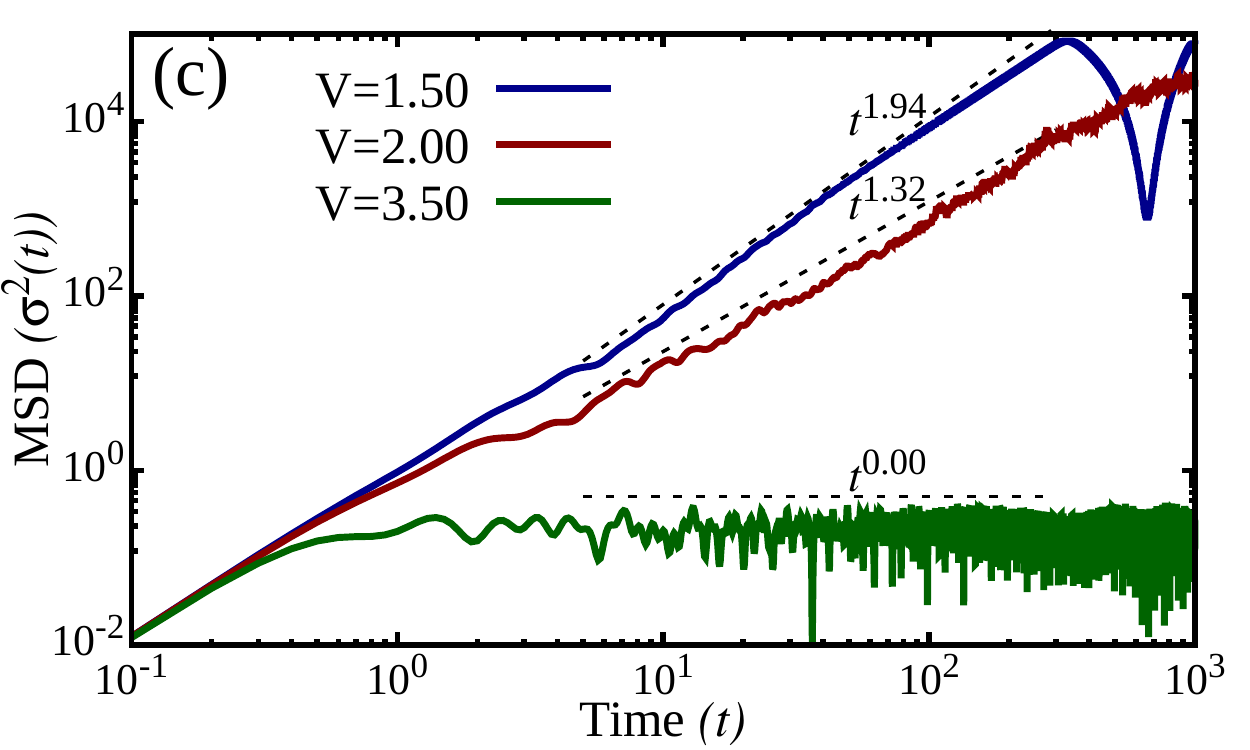} 	 \includegraphics[width=0.23\textwidth,height=0.22\textwidth]{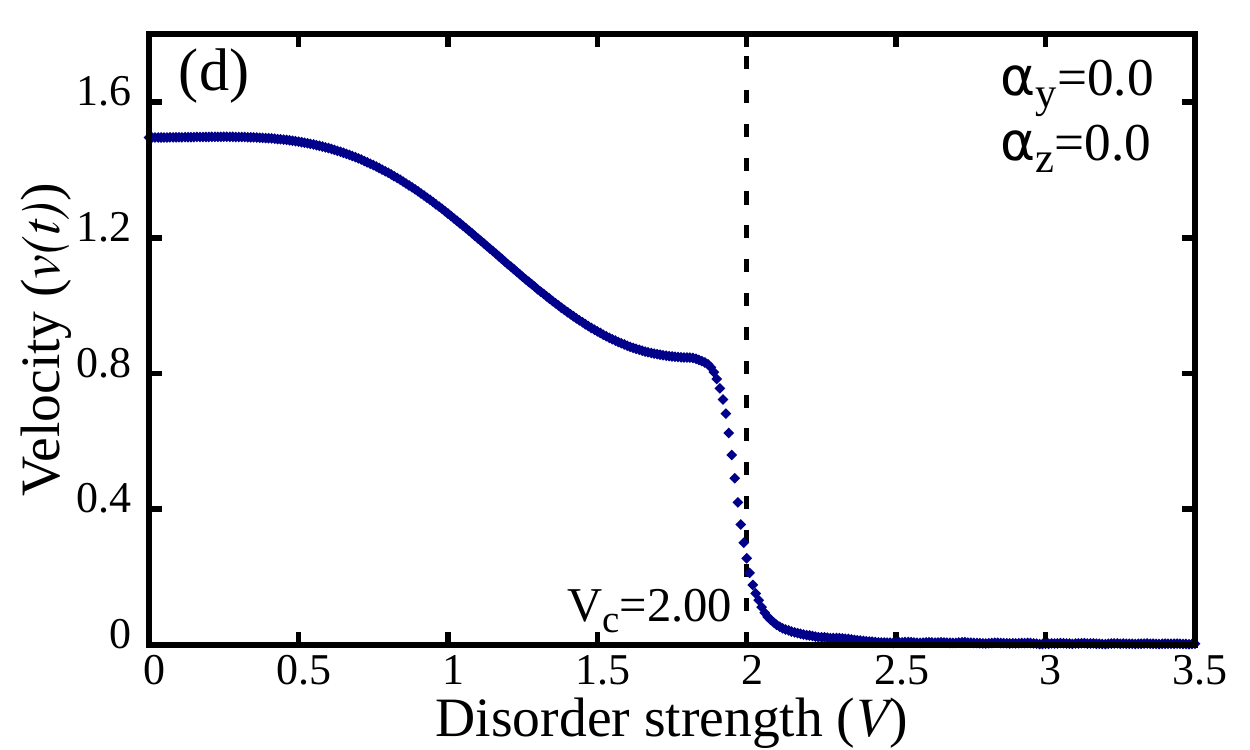}	
				
			\end{tabular}	
		\end{center}
		\vspace{-0.8cm}
		\caption{The Mean Square Displacement(MSD) as a function of time in double-logarithmic scale and its corresponding dynamical phase transition 
		(velocity $v(t)$ with an increasing strength of the quasiperiodic potential) for two cases: (i) Figs.~(a) and (b) are for a Hermitian system ($J_L=J_R=1$) with 
		RSO interaction ($\alpha_y=0.5$ and $\alpha_z=0.0$). (ii) Fig.~(c) and (d) are for a non-Hermitian system ($J_L/J_R=0.5$) without RSO interaction 
		($\alpha_y=0.0$ and $\alpha_z=0.0$). In the left panels, the dark-blue, dark-red and dark-green indicate the spectrally delocalized, critical and localized regimes 
		respectively. The values of $\mu$ (in Eq.~(\ref{Eq:MSD_Diffusion_exponent_relation})) have been determined by using linear fitting (indicated  by dashed lines).
		In the right panels, the vertical black line indicates the dynamical phase transition estimated at $t=150$ secs.}
		\label{Fig:MSD_Velocity_1}
	\end{figure}
	
	\begin{figure*}[]
		\centering 
		\includegraphics[width=0.24\textwidth,height=0.22\textwidth]{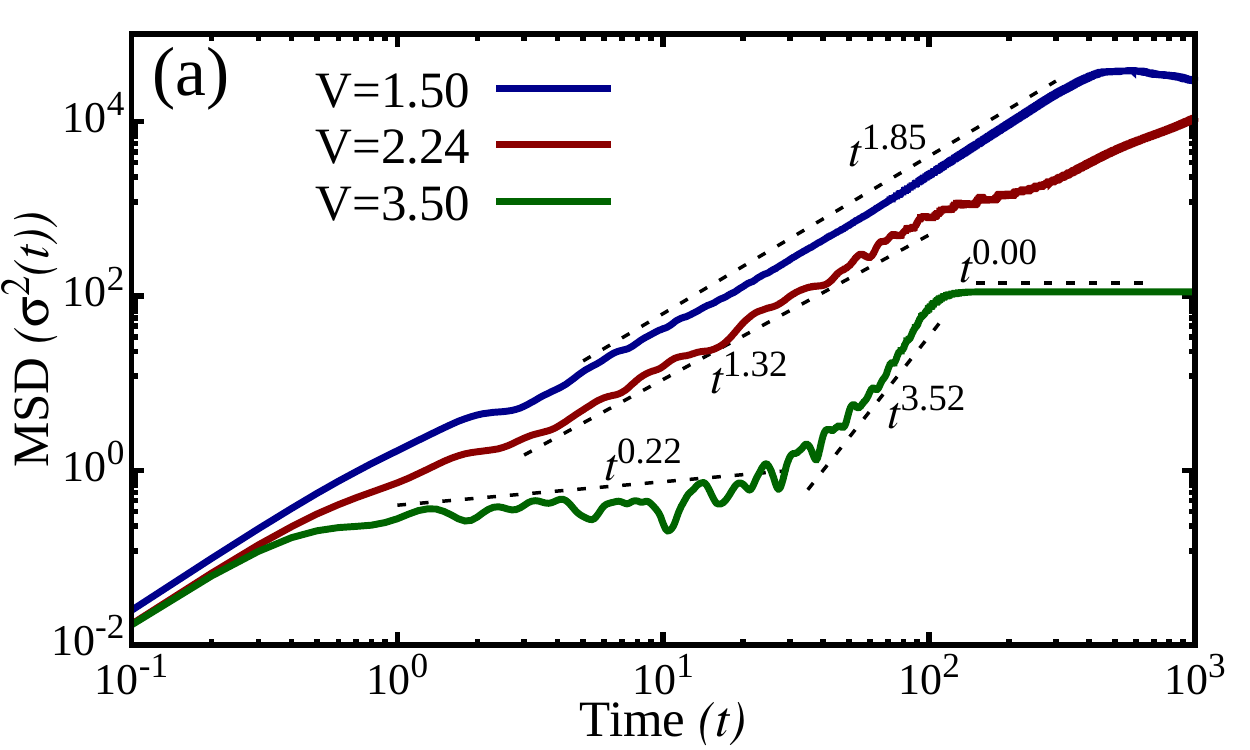}\hspace{0.0cm}
		\includegraphics[width=0.24\textwidth,height=0.22\textwidth]{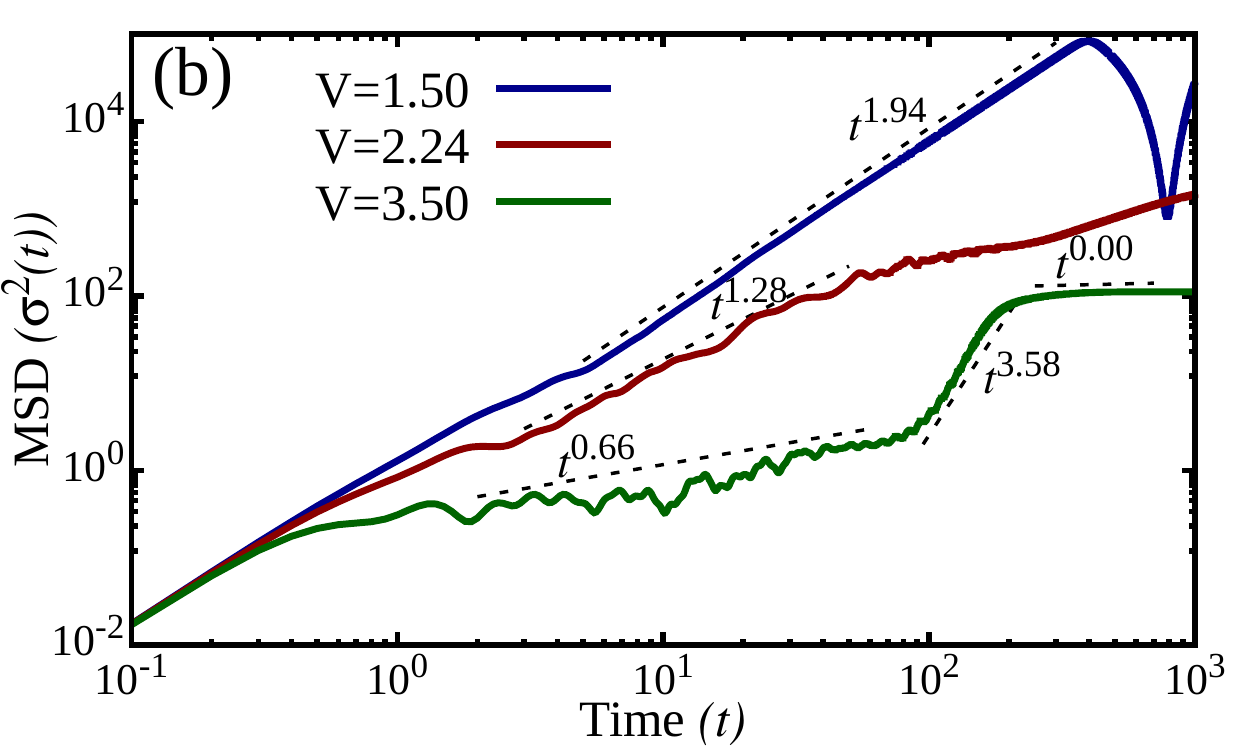}\hspace{0.0cm} 
		\includegraphics[width=0.24\textwidth,height=0.22\textwidth]{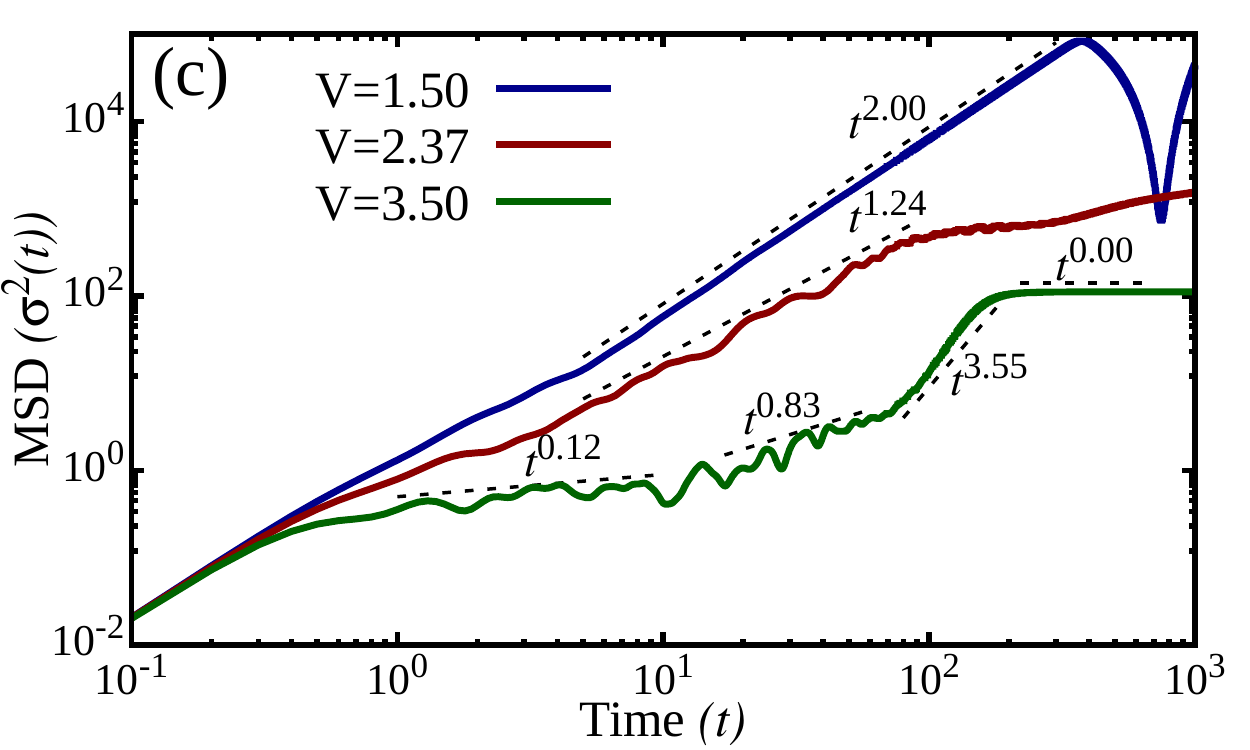}\hspace{0.0cm}
		\includegraphics[width=0.24\textwidth,height=0.22\textwidth]{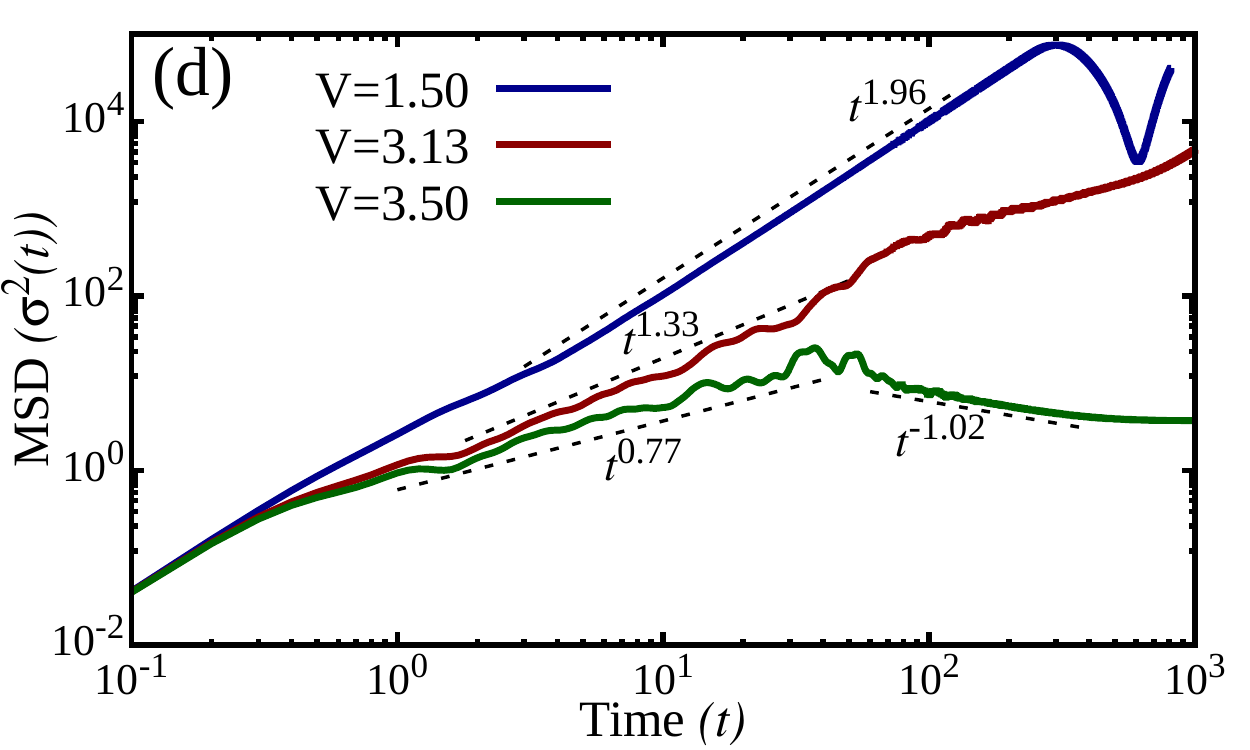}\\
		\includegraphics[width=0.24\textwidth,height=0.22\textwidth]{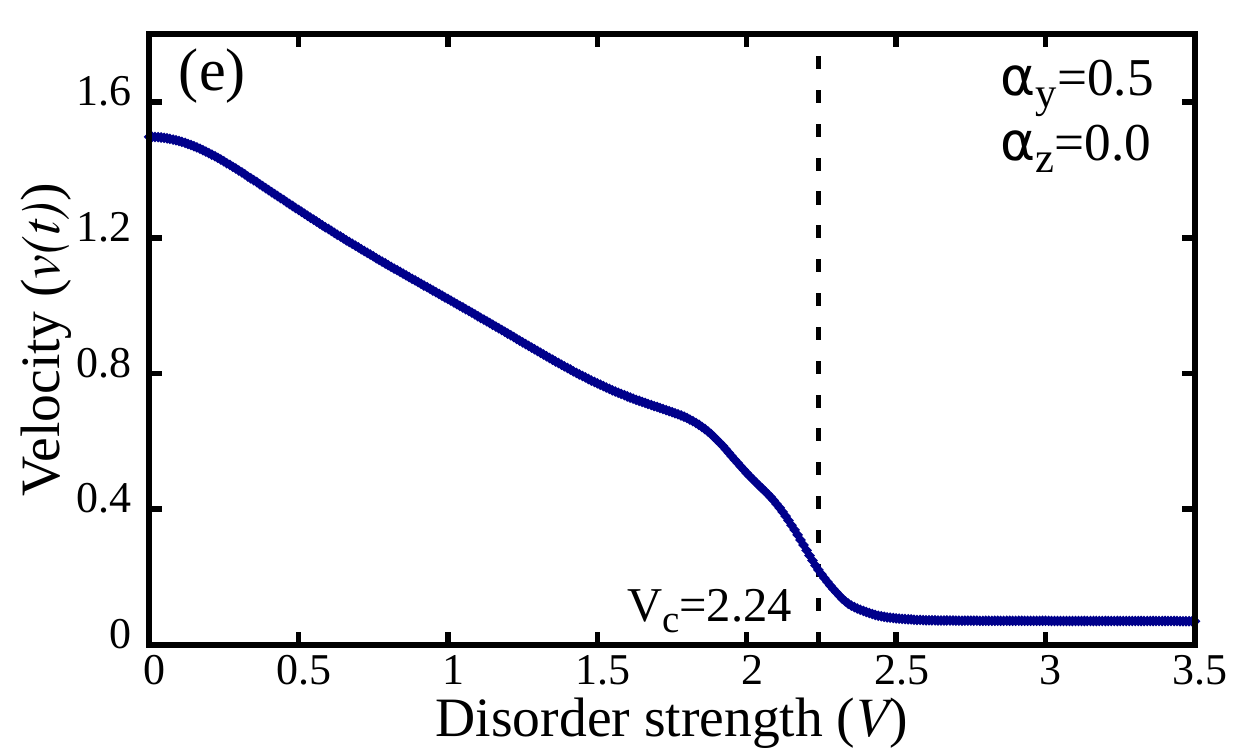}\hspace{0.0cm}
		\includegraphics[width=0.24\textwidth,height=0.22\textwidth]{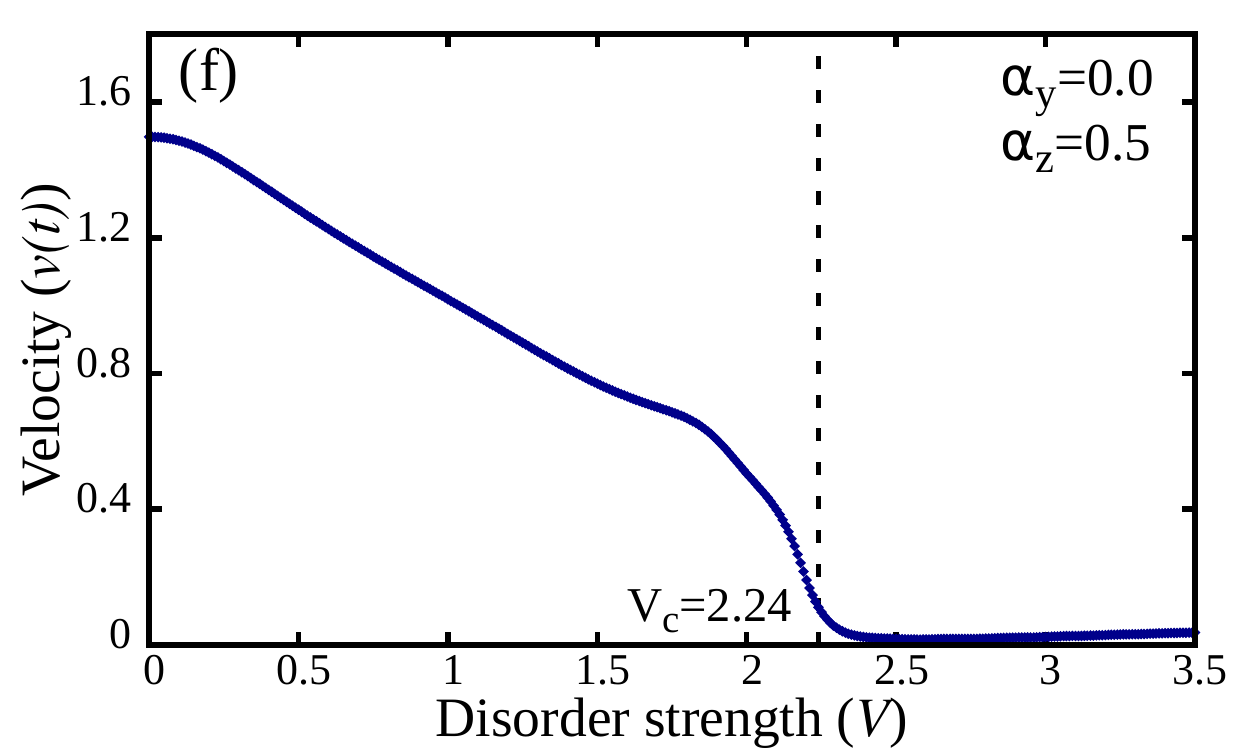}\hspace{0.0cm} 
		\includegraphics[width=0.24\textwidth,height=0.22\textwidth]{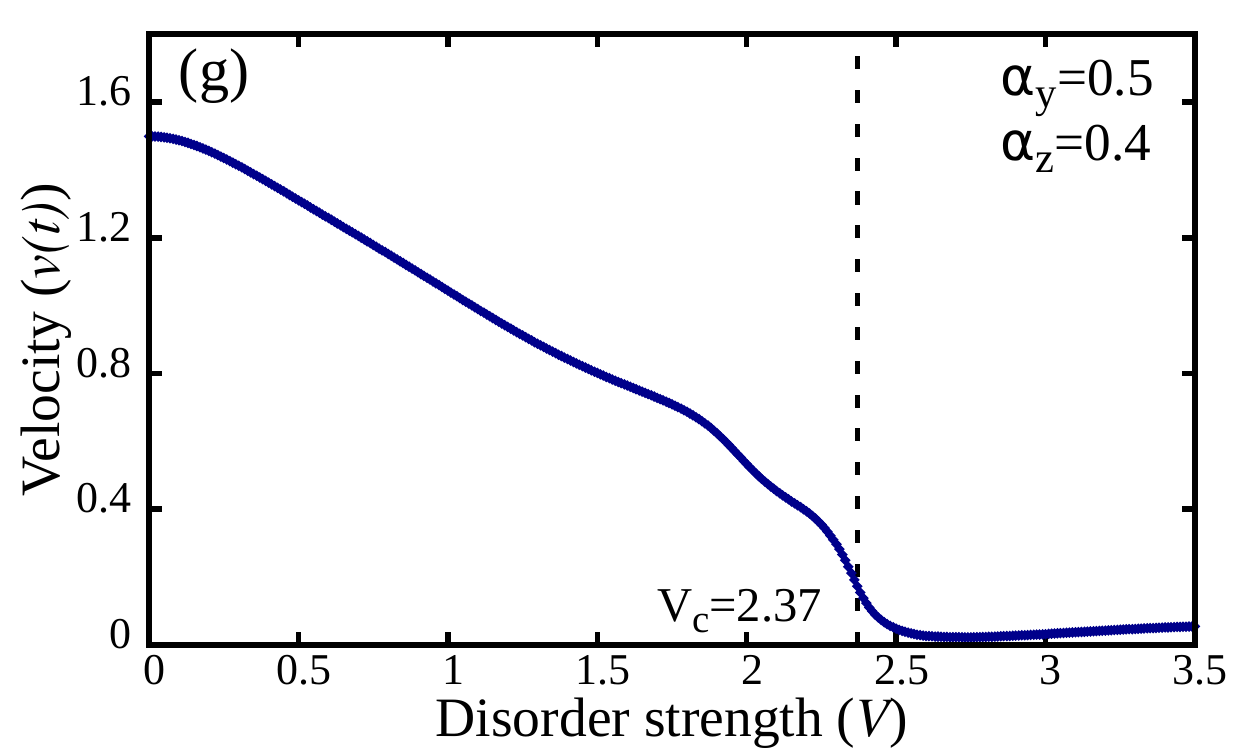}\hspace{0.0cm}
		\includegraphics[width=0.24\textwidth,height=0.22\textwidth]{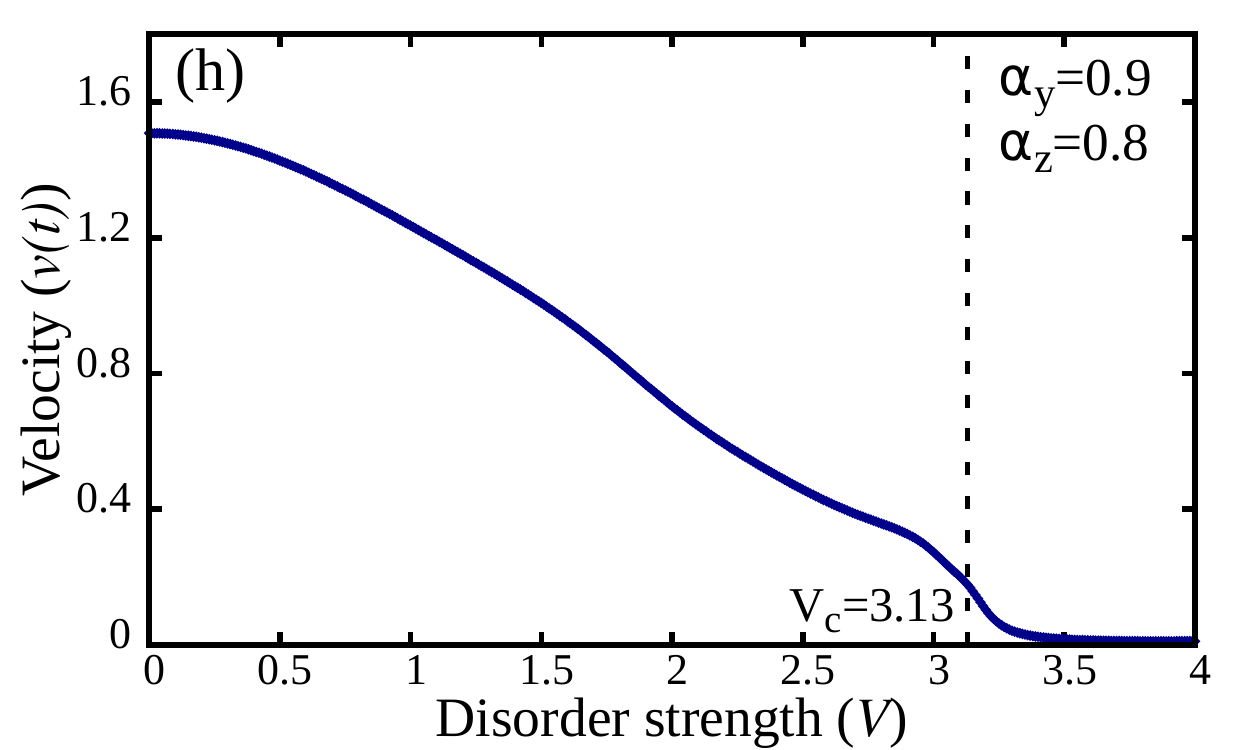}
		\vspace{-0.2cm}
		\caption{The MSD as a function of time in double-logarithmic scale in the upper panel (a-d) and its corresponding dynamical phase transition in the lower panel 
		(e-h) for the different non-Hermitian ($J_L/J_R=0.5$) cases: (i) Figs.~(a) and (e) are in the presence of only the spin-conserving hopping amplitude in the 
		RSO interaction ($\alpha_y=0.5$ and $\alpha_z=0.0$).
		(ii) Figs.~(b) and (f) are in the presence of only the spin-flip hopping amplitude in the RSO interaction ($\alpha_y=0.0$ and $\alpha_z=0.5$).
		(iii) Figs.~(c) and (g) are in the presence of both the spin-conserving and spin-flip hopping amplitudes in the RSO interaction ($\alpha_y=0.5$ and $\alpha_z=0.4$). 
		(iv) Figs.~(d) and (h) are in the presence of strong RSO interaction ($\alpha_y=0.9$ and $\alpha_z=0.8$). Similar to Fig.~(\ref{Fig:MSD_Velocity_1}), 
		in the upper panels, the dark-blue, dark-red and dark-green indicate the spectrally delocalized, critical and localized regimes respectively.
		In the bottom panels, the vertical black line indicates the dynamical phase transition calculated at $t=150$ secs.}
		\label{Fig:MSD_Velocity_2}
	\end{figure*}

	\section{Dynamics of an initially localized wave-packet}\label{Sec:Dynamical_study}
	From the discussions of the previous section it is evident that the RSO interaction introduces non-trivial spectral features in non-Hermitian systems 
	with asymmetric hopping.
	It is then natural to ask whether the RSO interaction affects the dynamical attributes of such systems at long times. To address this issue, we
	study the dynamics of a spin-up electron initially localized around the centre of the lattice $n_0$, under PBC. In Appendix~\ref{App:MSD_NH}, we have verified 
	that although the spectral features are sensitive to the boundary conditions, the dynamics of such an excitation is independent of it. 
	To generate the desired excitation, we consider 
	a single-particle basis and choose the Wannier state $\ket{n,\sigma} = 
	\ket{...0_\uparrow,0_\uparrow,1_\uparrow,0_\uparrow,0_\uparrow...,...0_\downarrow,0_\downarrow,0_\downarrow,0_\downarrow,0_\downarrow...}$. 
	The excitation of the wave-packet is thus a delta function $\delta_{n,n_0}$ in the up-spin channel, concentrated at the centre.
	The wave-packet is released at time $t=0$ and its dynamics is governed by Eqs.~(\ref{Eq:Spin_up_eqn}) and (\ref{Eq:Spin_down_eqn}). The 
	time-evolved Bloch states are obtained by superposing the evolved coefficients \cite{Katsanos}, and is given by,
	\begin{eqnarray}
		\psi(t)=\displaystyle\sum_{n,\sigma} \exp(-iE_n^{\sigma}t/\hbar) a_{n}^{\sigma}(0)\psi_n^{\sigma}(0)
		\label{Time-evolved_state}
 	\end{eqnarray}
	where, $a_n^{\sigma}(0)$'s are the initial coefficients of the wave-packet $\psi_{n}^{\sigma}(t=0)$.\\
	\indent It is well-known that the non-Hermitian systems exchange energy with the environment, leading to a non-unitary dynamics.
	Such systems also violate the conservation of probability, and the norm expands/shrinks with time.
	The evolution of the wavefunction after an interval $dt$ is thus achieved by a two-step process,
	\begin{eqnarray}
	 \ket{\psi(t+dt)}=\exp(-i\mathcal{H}dt/\hbar)\ket{\psi(t)},\nonumber
	\end{eqnarray}
	followed by,
	\begin{eqnarray}
	\ket{\psi(t+dt)}=\frac{\ket{\psi(t+dt)}}{||\ket{\psi(t+dt)}||}.
	\label{Eq:Time-evolved_norm_state}
	\end{eqnarray}
	To gain some qualitative insights of the dynamics, we present the amplitude of the time-evolved state ($\vert\psi_n(t)\vert$) 
	for different lattice sites as a function of time in Fig.~(\ref{Fig:Wave_profile}).
	Figs.~\ref{Fig:Wave_profile}(a-c) illustrate the change in the wave-profile with symmetric (Hermitian) hopping and RSO interaction $(\alpha_y=0.5, \alpha_z=0)$, 
	in the order of increasing strengths of the quasiperiodic potential. In this case, there is a DL transition at a critical value of $V_c \simeq 2.24,$ as determined in 
	Appendix \ref{App:IPR}. It is important to note that the qualitative features of the dynamics are identical to a Hermitian system without the RSO interaction, where 
	$V_c = 2.0$.
	From Fig.~\ref{Fig:Wave_profile}(a), it is clear that the excitation propagates throughout the lattice even at a long time, as expected in the spectrally 
	delocalized regime.
	On the other hand, from Fig.~\ref{Fig:Wave_profile}(c) it is evident that the wave-packet is dynamically localized at the site where it was 
	initially released ($n_0$), concurrent with the spectral localization. 
	Fig.~\ref{Fig:Wave_profile}(b) is an indicative of the multifractal character in the critical regime \cite{S_Chen}.
	We have verified that the interchange in the magnitudes of the RSO interaction ($\alpha_y=0.0$, $\alpha_z=0.5$) neither changes the critical value ($V_c$), nor the qualitative dynamical features.
	Moreover, such overall dynamical characteristics remain unaltered in the presence of both the spin-conserving and spin-flip amplitudes of the RSO 
	interaction ($\alpha_y=0.5$, $\alpha_z=0.4$), except that the critical point is pushed to a higher strength ($V_c\simeq2.37$) of the quasiperiodic
	potential as already discussed in Sec.~\ref{Sec:IPR}.
	It is thus primarily established that the addition of RSO interaction in such Hermitian systems induces no qualitative change in the wave-packet evolution.

	To understand the dynamical propagation of the excitation in a non-Hermitian system, we begin by switching the RSO interaction strengths to zero, 
	demonstrated in Figs.~\ref{Fig:Wave_profile}(d-f).
	Figs.~\ref{Fig:Wave_profile}(d) and (f) convey similar qualitative attributes below and above the critical point ($V_c=2.0$) as in Figs.~\ref{Fig:Wave_profile}(a) and (c),
	except the uni-directional light-cone nature of the wave evolution due to the asymmetric hopping ($J_L/J_R=0.5$). The preferential one-way propagation arises because of 
	the unbalanced amplification/attenuation of the counter-propagating wave-packets \cite{Longhi2015}. Although, it is evident that the non-Hermitian systems 
	without RSO interaction undergoes a dynamical transition accompanying the spectral transition similar to its Hermitian counterpart, the nature of the 
	wave-packet spreading at the critical point are not identical as can be seen from Figs.~\ref{Fig:Wave_profile}(b) and (e).
	
	The third panel of Fig.~(\ref{Fig:Wave_profile}) depicts the evolution of the wave-packet in non-Hermitian systems for different strengths of the RSO interaction.
	In stark contrast to Figs.~\ref{Fig:Wave_profile}(c) and (f) in the spectrally localized regime, 
	we observe that the wave slides to a nearby lattice site ($n=315$) in Fig.~\ref{Fig:Wave_profile}(g)
	when either the spin-conserving or the spin-flip hopping amplitudes of the RSO interaction are non-zero, i.e., $\alpha_y(\alpha_z)$=0.5 and $\alpha_z(\alpha_y)$=0.0, 
	where $V_c\simeq2.24$. Furthermore, as evident from Fig.~\ref{Fig:Wave_profile}(h), this feature remains intact beyond the critical point ($V_c\simeq2.37$)
	in such non-Hermitian systems with both the components of the RSO interaction being non-zero ($\alpha_y=0.5$, $\alpha_z=0.4$).
	Interestingly, however, in both the above cases, the excitation possesses some tendency to propagate even beyond the critical thresholds in the spectrally localized regime.
	Such a drift from the initial site of excitation in the spectrally localized regime is in contrast to the behavior of being completely localized at $n_0$ as observed
	in Figs.~\ref{Fig:Wave_profile}(c) and (f). This characteristic is typical of non-Hermitian systems, and is termed as non-Hermitian jump \cite{Tzortzakakis, S_Chen}.
	Moreover, Fig.~\ref{Fig:Wave_profile}(i) clearly suggests the absence of complete localization even at long time
	in the spectrally localized regime (beyond $V_c\simeq3.13$) when the RSO interaction strengths are sufficiently strong enough ($\alpha_y=0.9, \alpha_z=0.8$).
	From these discussions, it is evident that the interplay between the non-Hermiticity and the RSO interaction renders unusual dynamical properties, 
	which requires further quantitative analysis, and will be dealt with in the forthcoming sections.
	\subsection{The Mean Square Displacement, diffusivity, and spreading velocity}\label{Sec:MSD_and_diffusivity}
	
	To assess the dynamics at a long time, we begin with the general definition of the Mean Square Displacement (MSD)
	for the two spin-orientations given as \cite{Zhang,Dai,Longhi,S_Chen},
	\begin{eqnarray}
		\sigma^2(t)=\frac{\sum_{n,\sigma} (n-n_0)^2|\psi_n^{\sigma}(t)|^2}{\sum_{n,\sigma} |\psi_n^{\sigma}(t)|^2}
		\label{Eq:Mean_Square_Displacement}
	\end{eqnarray}
	Eq.~(\ref{Eq:Mean_Square_Displacement}) gives the positional variance of the wave-packet, and describes the temporal spreading around the initially excited lattice site $n_0$.
	The asymptotic dependence of MSD upon time is given as \cite{Katsanos},
	\begin{eqnarray}
		\sigma^2(t)=t^{2\delta}\simeq t^{\mu},
		\label{Eq:MSD_Diffusion_exponent_relation}
	\end{eqnarray}
	where $\delta=\mu/2$ is identified as the diffusion exponent and describes the nature of wave-packet spreading.
	The electron transport is termed as ballistic when $\mu\simeq2$, whereas it is diffusive when $\mu\simeq1$.
	Moreover, the transport may be sub-diffusive, super-diffusive and hyper-diffusive when $\mu<1$, $1<\mu<2$ and $\mu>2$ respectively.
	Typically, it is expected that when the strength of the quasiperiodic potential exceeds the critical value($V>V_c$), the wave-packet is confined 
	at its initial site of release, giving rise to complete dynamical localization for all times ($\mu=0$).
	Furthermore, as a general rule, the diffusion takes place from the region of high concentration of the mobile charge carriers to a region of a comparatively lower concentration.\\
	\indent
	In Ref.~\cite{Longhi}, the author has introduced spreading velocity as a measure to determine the dynamical phase transition. 
	The spreading velocity of the initial excitation is defined as,
	\begin{eqnarray}
		v(t)\sim\frac{\sigma(t)}{t}.
		\label{Eq:Spreading_velocity}
	\end{eqnarray}
	A non-vanishing speed indicates a ballistic motion, whereas in the absence of transport, $v(V)=0$ for all times.
	At the dynamical phase transition ($V=V_c$), the transport is intermediate between the completely localized and ballistic regimes ($\mu\simeq1$).
	Such dynamical measures often find applications in photonic lattices \cite{Lahini,Iomin,Xu}. 

	In Figs.~\ref{Fig:MSD_Velocity_1}(a-b), we have demonstrated the behavior of the MSD with time and its corresponding dynamical phase transition 
	for a Hermitian system with RSO interaction, where we consider the spreading velocity $v(V)$ as an order parameter.
	The behavior of the MSD in Fig.~\ref{Fig:MSD_Velocity_1}(a) clearly indicates the anticipated values of the diffusion exponent, i.e.- $\delta(=\mu/2)=0.94, 0.48$ and $0$ 
	for the delocalized, critical and localized regimes respectively. The speed of propagation in Fig.~\ref{Fig:MSD_Velocity_1}(b) decreases continuously with $V$,
	and vanishes in the localized regime ($V>V_c$) due to the suppression of the wave-packet spreading after the dynamical phase transition, that indicates a second order phase transition. 
	Figs.~\ref{Fig:MSD_Velocity_1}(c-d) demonstrates the behaviour of a non-Hermitian system ($J_L/J_R=0.5$) in the absence of RSO interaction.
	The diffusive exponent as shown in Fig.~\ref{Fig:MSD_Velocity_1}(c) exhibits an almost similar behavior as in Fig.~\ref{Fig:MSD_Velocity_1}(a), except 
	that the wave-packet spreading in the critical regime is super-diffusive with $\delta=0.66$. This finding is consistent with the observations 
	of the wave profile discussed in the preceding section. However, the discontinuity in Fig.~\ref{Fig:MSD_Velocity_1}(d) is suggestive of a first-order phase 
	transition and has been recently reported in Ref.~\cite{Longhi}. It is important to note that in both the scenarios, the dynamical phase transition coincides 
	with the spectral phase transition and follows the general expectation, which is demonstrated in Appendix~\ref{App:IPR}.
	
		\begin{figure*}[]
		\centering 
		\includegraphics[width=0.24\textwidth,height=0.23\textwidth]{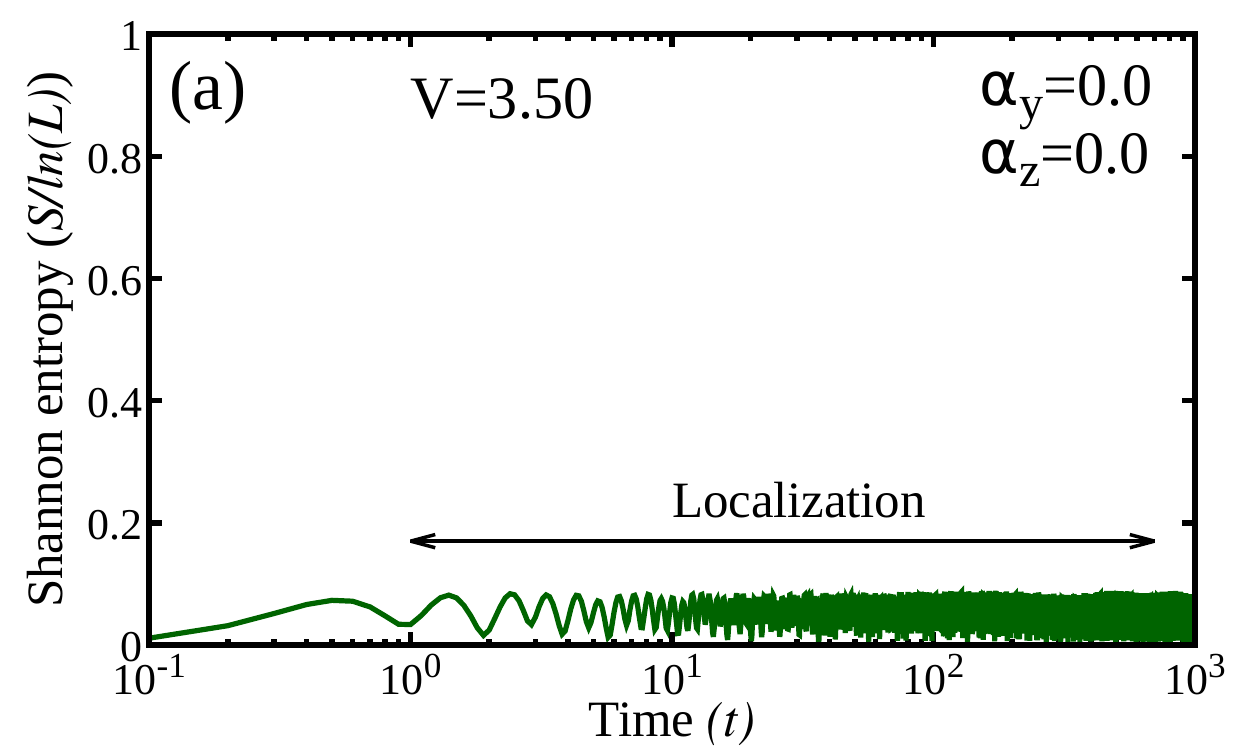}
		\includegraphics[width=0.24\textwidth,height=0.23\textwidth]{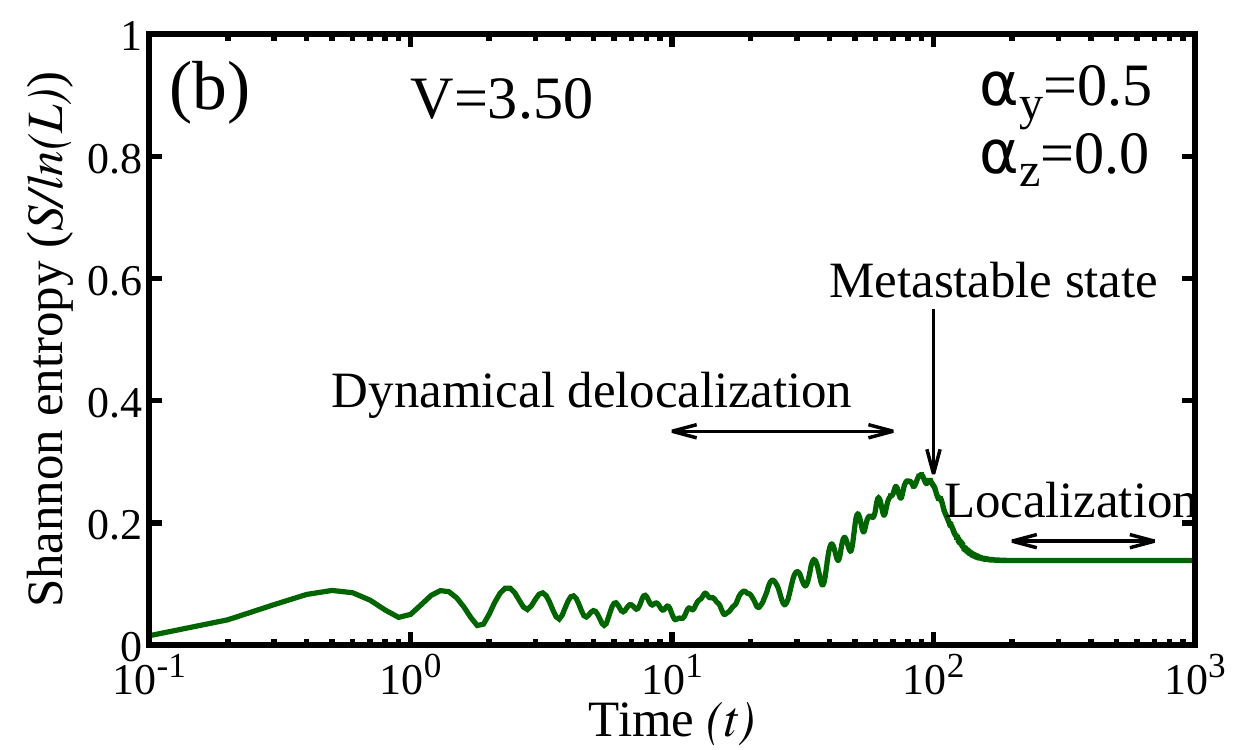}
		\includegraphics[width=0.24\textwidth,height=0.23\textwidth]{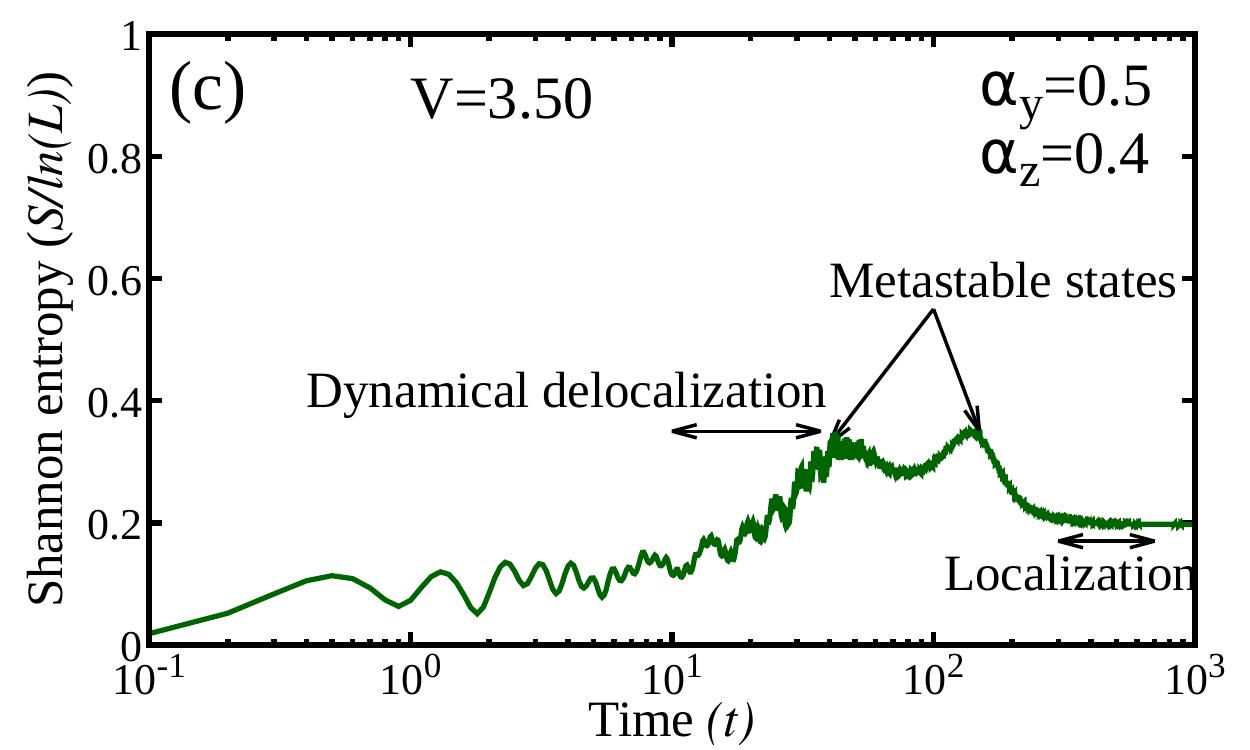}
		\includegraphics[width=0.24\textwidth,height=0.23\textwidth]{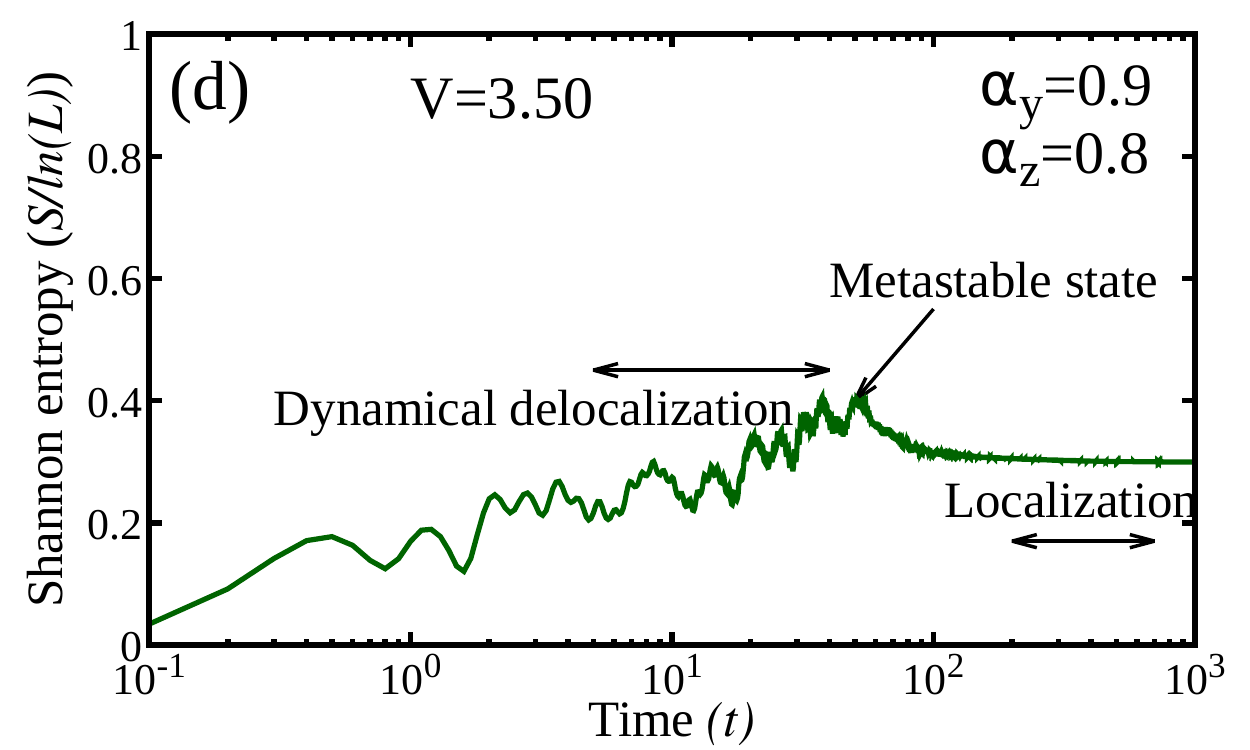}
		\caption{The behavior of Shannon entropy ($S/ ln(L)$) with time for a $610$-site system. The dynamical delocalization in the spectrally localized regime ($V$=3.50) along with the metastable states is demonstrated for a non-Hermitian system ($J_L/J_R=0.5$) for four distinct cases: (a) absence of RSO interaction ($\alpha_y=0.0$ and  $\alpha_z=0.0$), (b) in the presence of only the spin-conserving hopping ($\alpha_y=0.5$ and $\alpha_z=0.0$), (c) in the presence of both the spin-conserving and spin-flip hopping ($\alpha_y=0.5$ and $\alpha_z=0.4$), and (d) for strong RSO interaction where negative diffusion occurs ($\alpha_y=0.9$ and $\alpha_z=0.8$).}
		\label{Fig:Shannon_entropy}
	\end{figure*}

	\indent
	However, this anticipated behavior changes significantly when we consider the non-Hermitian system in the presence of RSO interaction.
	Fig.~\ref{Fig:MSD_Velocity_2} illustrates the non-trivial behavior of our asymmetric non-Hermitian Hamiltonian in the presence of RSO interaction.
	Figs.~\ref{Fig:MSD_Velocity_2}(a,e) demonstrates the behavior of the dynamical attributes when the spin-conserving RSO interaction amplitude is non-zero.
	From Fig.~\ref{Fig:MSD_Velocity_2}(a), it is evident that when $\alpha_y=0.5$, the initially excited spin-up electron has a tendency to diffuse even in the 
	spectrally localized regime (as shown in dark-green line). Moreover, the behavior is hyper-diffusive for $t\simeq20-100$ secs ($\delta=1.76$). However, such a 
	diffusion is confined only to the nearby lattice sites as discussed in the initial results of Sec.~\ref{Sec:Dynamical_study}. Thereafter, the wave-packet is 
	observed to localize at all times. We have verified the absence of finite size effect on the MSD estimates in Appendix~\ref{App:MSD_NH}.\\
	\indent
	Figs.~\ref{Fig:MSD_Velocity_2}(b,f) illustrate the dynamical behavior of the wave-packet in the presence of only the spin-flip hopping amplitude ($\alpha_z=0.5$).
	It is discernible that the presence of either the spin-conserving or spin-flip hopping amplitude of the RSO interaction renders dynamical delocalization with similar traits,
	except that the localization occurs after a longer time ($t\simeq220$ secs) in the latter case.
	Figs.~\ref{Fig:MSD_Velocity_2}(c,g) convey the same observations when both $\alpha_y$ and $\alpha_z$ in the RSO interaction are non-vanishing, 
	except that in this case there is an additional sub-diffusive regime before the hyper-diffusive transport sets in. 
	A similar tendency of hyper-diffusivity in the localized regime has been pointed out in a recent work \cite{S_Chen}.
	In addition, such a dynamical delocalization has been realized experimentally \cite{Weidemann}. 
	However, the authors of Ref.~\cite{S_Chen} have attributed the hyper-diffusive transport
	to the breaking of $\mathcal{PT}$ symmetry and attributed its existence only in the broken $\mathcal{PT}$-symmetric regime.
	In contrast, in our non-Hermitian system (which is intrinsically in the broken $\mathcal{PT}$-symmetric regime due to asymmetric
	hopping amplitudes), we have illustrated in Fig.~\ref{Fig:MSD_Velocity_1}(c), that this hyper-diffusive transport does not appear even when the 
	$\mathcal{PT}$-symmetry is broken. On the other hand, it is clear from Fig.~\ref{Fig:MSD_Velocity_1}(a), that the dynamical delocalization is absent in 
	Hermitian systems even in the presence of RSO interaction. Hence, we can infer that in our non-Hermitian Hamiltonian the unusual dynamical 
	features in the spectrally loclalized regime appears due to the combined effects of non-Hermiticity and the RSO interaction, as observed
	in Figs.~\ref{Fig:MSD_Velocity_2}(a-c). A discussion of the MSD calculations with varying $J_L/J_R$ ratios governing the non-Hermiticity has been 
	relegated to Appendix~\ref{App:MSD}.\\

	\indent
	We find an even more surprising dynamical feature when we consider the non-Hermitian system with strong RSO interaction amplitudes, where $\alpha_y=0.9$ and $\alpha_z=0.8$. 
	It is clear from Fig.~\ref{Fig:MSD_Velocity_2}(d) that a sufficiently strong RSO interaction can cause the transport
	from a low concentration to a higher concentration, and might result in a negative diffusivity ($\delta\simeq-0.50$) instead of a hyper diffusive 
	behavior before dynamical localization. 
	The negative diffusion is uncommon and has been reported in a two-dimensional lattice-gas system with attractive interaction \cite{Argyrakis}.
	 
	\subsection{Behavior of the Shannon entropy in the non-Hermitian spectrally localized regime}\label{Sec:Shannon_entropy}
	To further identify the dynamical delocalization properties in the localized regime using the properties of eigenstates at different times,
	we use the measure of Shannon entropy for the time-evolved states $\psi_n(t)$ defined as \cite{Santos1,Santos2,Alessio,Sarkar},
	\begin{eqnarray}
		S(t)=-\displaystyle\sum_{n,\sigma} |\psi_n^{\sigma}(t)|^2 \ln |\psi_n^{\sigma}(t)|^2,
	\label{Eq:Shannon_entropy}
	\end{eqnarray}	
	where $|\psi_n(t)|$ is the amplitude of the eigenstate at the $n^{th}$ site and at time $t$.
	It is well known that the Shannon entropy for localized states $S(t)\rightarrow0$, whereas delocalization leads to maximum entropy (due to the randomness 
	generated in the itinerant behavior),
	$S(t)\simeq\ln L$ (since $|\psi_n(t)| \propto 1/\sqrt{L})$.
	A plot of $S/\ln L$ as a function of time (in Fig.~(\ref{Fig:Shannon_entropy_Hermitian}) of Appendix~\ref{App:Hermitian_Shannon_entropy}) indicates a change as shown for the Hermitian case with RSO interaction (corresponding to Fig.~\ref{Fig:MSD_Velocity_1}(a)).
	As expected, $S/\ln L\sim O(1)$ in the delocalized regime at long times, whereas,
	$S/\ln L$ remains a constant (approaching zero) as soon as $V$ crosses its critical threshold at $V_c$.
	The behavior in the critical regime is intermediate between the delocalized and localized phases as evident from Fig.~(\ref{Fig:Shannon_entropy_Hermitian}).
	It is important to note that unlike the isolated systems, the entropy of a non-Hermitian system might decrease with time
	owing to the interaction with the environment. However, at equilibrium the total entropy of the system and its surroundings in such cases 
	remains unaltered \cite{HatanoTR}.\\
	\indent
	To obtain an inference of the stability of the dynamically delocalized eigenstates as discussed in Sec.~\ref{Sec:MSD_and_diffusivity}, we have compared the different cases of the non-Hermitian system with and without RSO interaction in the spectrally localized regime ($V=3.50$).
	In the absence of RSO interaction, Fig.~\ref{Fig:Shannon_entropy}(a) clearly portrays the dynamically localized regime, where the Shannon entropy remains at a minimum (nearly zero) upto long times. 
	Such a behavior is a clear manifestation of the dynamical equilibrium.
	However, in the presence of the spin-preserving RSO interaction ($\alpha_y=0.5$ and $\alpha_z=0.0$), the entropy initially increases, before reaching
	an intermediate metastable state (characterized by the maximized entropy) which exists for a short time, and finally decreases to reach the ultimate equilibrium after the dynamical localization (Fig.~\ref{Fig:Shannon_entropy}(b)). 
	Fig.~\ref{Fig:Shannon_entropy}(c) illustrates the behavior of the Shannon entropy in the presence of both
	the RSO interaction terms ($\alpha_y=0.5$ and $\alpha_z=0.4$). The non-Hermitian system behaves similar to Fig.~\ref{Fig:Shannon_entropy}(b) by attaining two intermediate metastable states (as is also evident from Fig.~\ref{Fig:MSD_Velocity_2}(c)), before reaching the dynamical equilibrium.
	However, in contrast to an ordinary diffusion where the entropy increases with time, in the presence
	of strong RSO interaction, Fig.~\ref{Fig:Shannon_entropy}(d) suggests that in the negative diffusion (where the transport is from a lower to a higher concentration), the entropy decreases with time and then reaches its final equilibrium.
	In addition, from Fig.~(\ref{Fig:Shannon_entropy}), it can be noted that the entropy at the ultimate equilibrium increases with an increase in the strength of the 
	RSO interaction. 
	
	\section{Conclusions}\label{Sec:Conclusions}
	In conclusions, this work demonstrates the unique spectral and dynamical properties in non-Hermitian quasicrystals with RSO interaction.
	An analysis of the energy spectrum of the non-Hermitian quasicrystals in the presence of RSO interaction
	clearly indicates that the states remain in the broken time-reversal symmetry for all strengths of the quasi-periodic potential,
	unlike in the cases without RSO interaction, which exhibits a real-complex energy transition alongwith the DL transition.
	We find that the association of the reality in the energy spectrum to the skin-effect is indeed a system-size dependent behavior.
	It is verified that the skin-modes can exist even in the complex spectral regime under an open boundary, in contrast to the previously reported results.
	Furthermore, it is demonstrated that when the magnitude of spin-flip term exceeds the magnitude of the spin-conserving term in the RSO interaction, the directionality of the skin-effect significantly reduces.
	The spin-dependent transport of the electrons has been thoroughly investigated using the diffusion exponent, determined from the variance of the wave-packet initially excited for the up-spin at the centre of the lattice.
	We have studied the dynamical phase transitions using the spreading velocity $v$. We have found evidences that although the spectral DL transition is concurrent to the dynamical phase transition, the interplay of the non-Hermiticity and the 
	RSO interactions leads to dynamical delocalization in the spectrally localized regime, in contrast to the Hermitian system with the RSO interaction and non-Hermitian 
	system without the interaction. 
	Interestingly, we have obtained hyper-diffusive behavior beyond the absolute spectral localization.
	In addition, we have illustrated that for strong RSO interaction, a negative diffusion from a lower to a higher concentration (as opposed to normal diffusive process) may be manifested in non-Hermitian systems with asymmetric hopping.
	In closing, we have shown that such non-Hermitian systems with RSO interaction goes through metastable state(s) before achieving dynamical equilibrium at a long time.

	\begin{acknowledgements} 
	A.C. acknowledges CSIR-HRDG, India, for providing financial support via File No.- 09/983(0047)/2020-EMR-I.
	The authors are thankful to the data computation facilities supported by SERB (DST), India (Grant No. EMR/2015/001227) and the High Performance Computing (HPC) facilities of the National Institute of Technology (Rourkela).
	\end{acknowledgements}

	\appendix
	
	\begin{figure}
		\begin{center}	
			\begin{tabular}{p{\linewidth}c}
				
				\includegraphics[width=0.24\textwidth,height=0.22\textwidth]{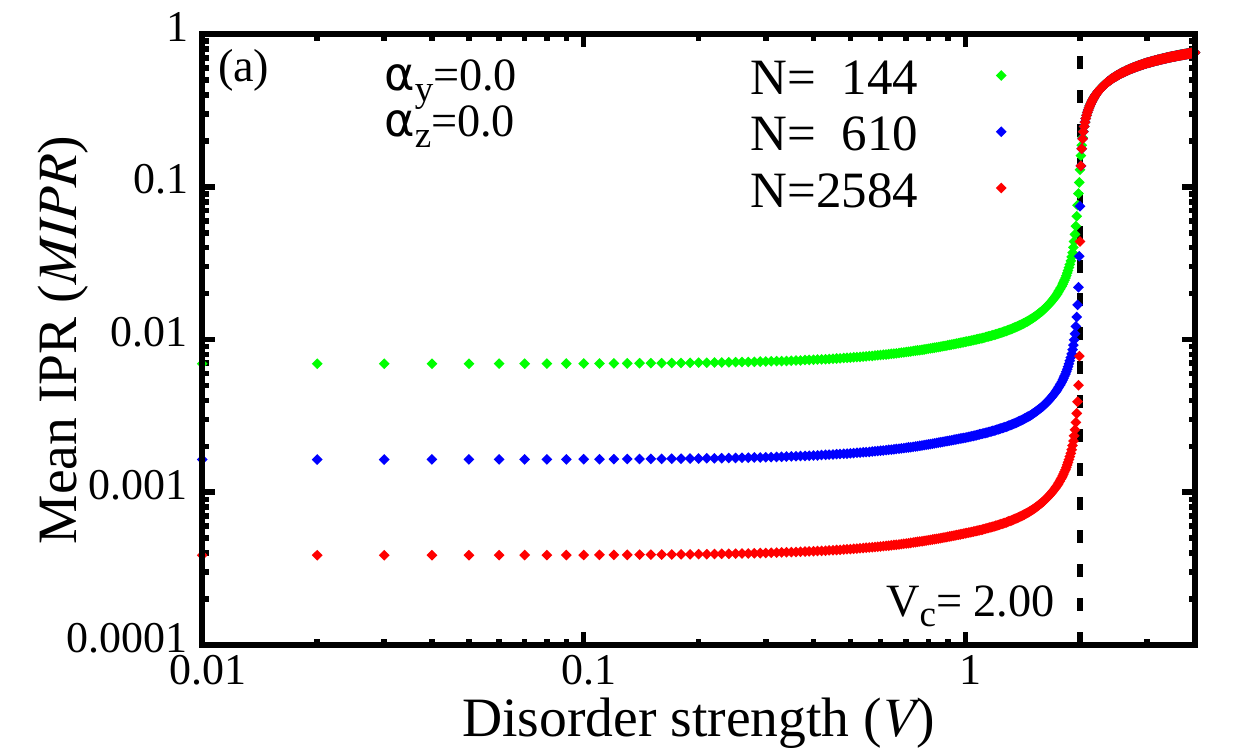} 	 \includegraphics[width=0.24\textwidth,height=0.22\textwidth]{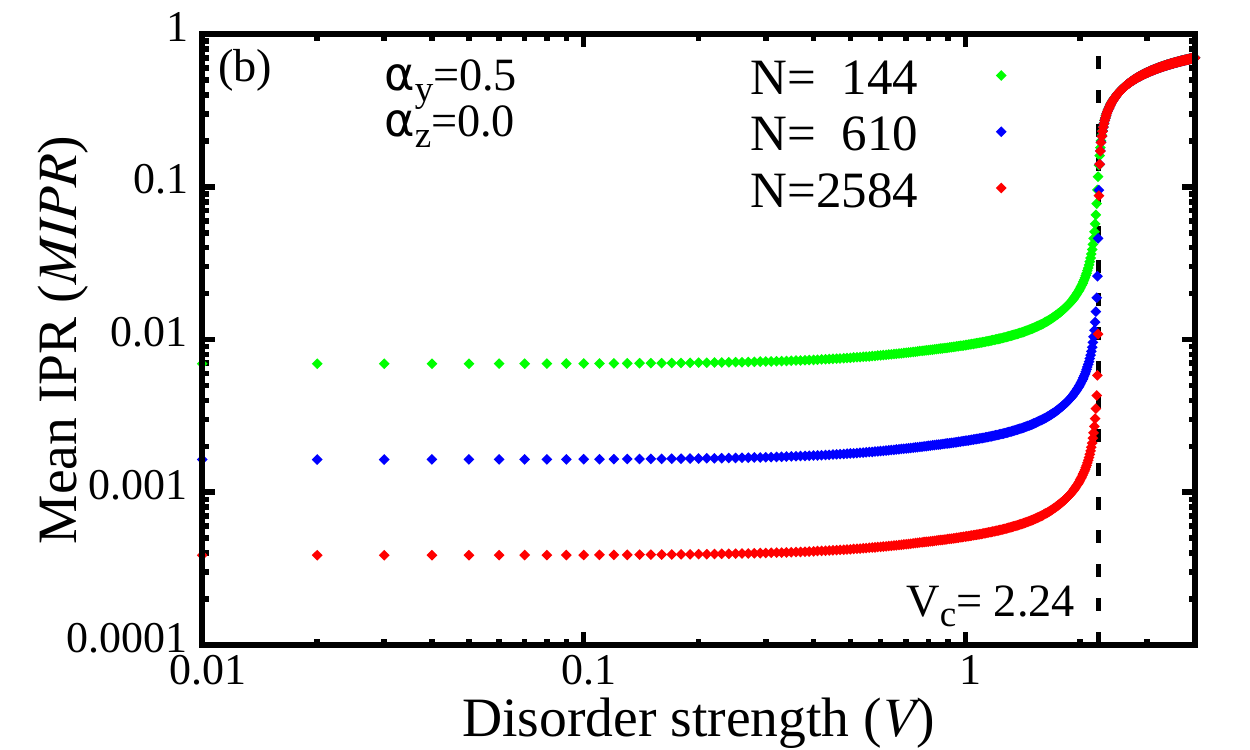}
				
			\end{tabular}
			
		\end{center}
		\vspace{-1cm}
		\begin{center}	
			\begin{tabular}{p{\linewidth}c}
				
				\includegraphics[width=0.24\textwidth,height=0.22\textwidth]{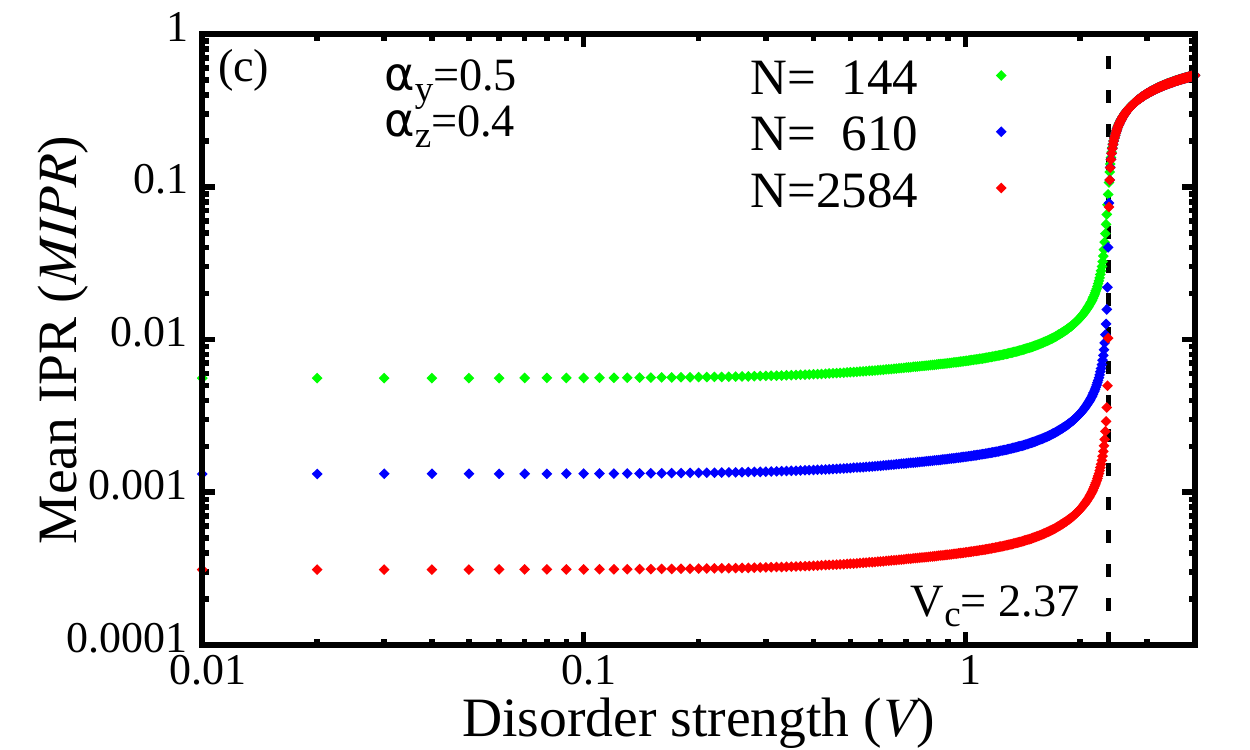} 	 \includegraphics[width=0.24\textwidth,height=0.22\textwidth]{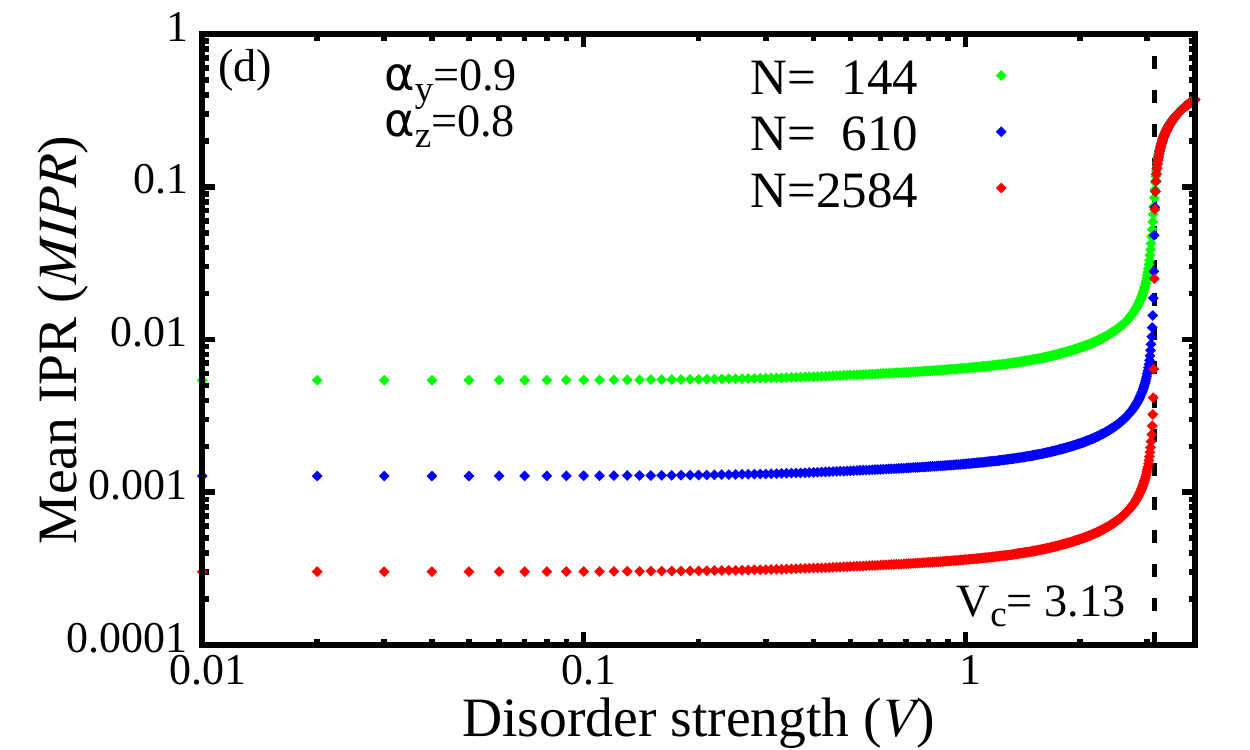}
				
			\end{tabular}	
		\end{center}
		\vspace{-0.8cm}
	    \caption{ The MIPR as a function of the quasiperiodic potential ($V$) for different values of $\alpha_y$ and $\alpha_z$ in the Hamiltonian with RSO interaction given by Eq.~(\ref{Eq:Total_Hamiltonian}). (a) $\alpha_y=0.0$ and $\alpha_z=0.0$, (b) $\alpha_y=0.5$ and $\alpha_z=0.0$, (c) $\alpha_y=0.5$ and $\alpha_z=0.4$, (d) $\alpha_y=0.9$ and $\alpha_z=0.8$. The critical point for the DL transition has been indicated by $V_c$ and represented by vertical black-dashed lines. The red, blue and green markers indicate the system sizes $L$=2584, 610 and 144 respectively. We have used the condition of non-Hermiticity, i.e.-$J_L/J_R=0.5$ and the periodic boundary condition in a lattice with $610$ sites in all the above cases.}
	    \label{Fig:MIPR}
    \end{figure}	

	\section{MIPR OF THE NON-HERMITIAN QUASI-PERIODIC HAMILTONIAN WITH AND WITHOUT RSO INTERACTION}\label{App:IPR}
	The authors of Ref.~\cite{Longhi} have indicated that the non-Hermitian ($J_L\neq J_R$ and $J_R>J_L$) phase transition occurs at the critical value $V_c$ given by,
	\begin{eqnarray}
		V_c=2 J_R,\nonumber 
	\end{eqnarray}
	which is also identical to the Hermitian case, when $J_L=J_R$.
	The most familiar approach used in extracting the critical point in DL transition is by determining the
	Inverse Participation Ratio(IPR) defined for a given state as \cite{Mirlin,Wessel},
	\begin{eqnarray}
		IPR_j=\frac{\displaystyle\sum_{n,\sigma} |\psi_{n,j}^{\sigma}|^4}{(\displaystyle\sum_{n,\sigma} |\psi_{n,j}^{\sigma}|^2)^2}
		\label{Eq: IPR}
	\end{eqnarray}
	The mean-IPR (MIPR) is then defined as \cite{S_Chen},
	\begin{eqnarray}
		MIPR=\frac{1}{N}\displaystyle\sum_{j}\frac{\sum_{n,\sigma} |\psi_{n,j}^{\sigma}|^4}{(\sum_{n,\sigma} |\psi_{n,j}^{\sigma}|^2)^2}
		\label{Eq: MIPR}
	\end{eqnarray}
	In the extended/delocalized regime, the MIPR varies inversely with the system size as $1/N$, and approaches zero in the thermodynamic limit.
	However, for the insulating/localized states, the IPR is independent of $N$ and approaches 1 with increasing strength of the disorder.\\
	\indent 
	Figs.~\ref{Fig:MIPR}(a-d) demonstrates the behavior of MIPR as a function of the strength of the disorder in double-logarithmic scale.
	The sharp jump in MIPR towards 1 clearly hints at the DL transition at that critical value of disorder, indicated by $V_c$.
	It is evident that $V_c$ increases with the strength of the RSO interaction.
	Hereafter, the dynamical study (in Sec.~\ref{Sec:Dynamical_study} of the main text) has been accomplished
	using the numerically determined values of $V_c$ from these numerical estimates.
	    
	    \begin{figure}
	    			
	    			\includegraphics[width=0.22\textwidth,height=0.20\textwidth]{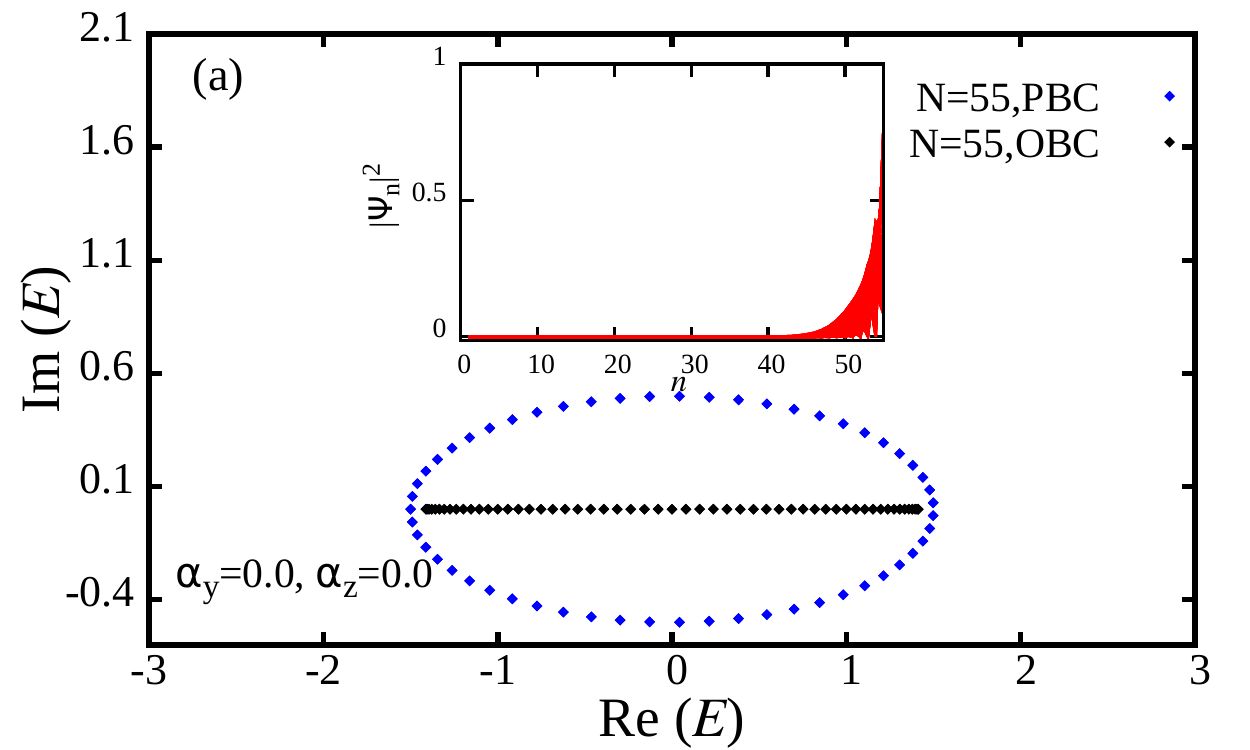} 	 \includegraphics[width=0.22\textwidth,height=0.20\textwidth]{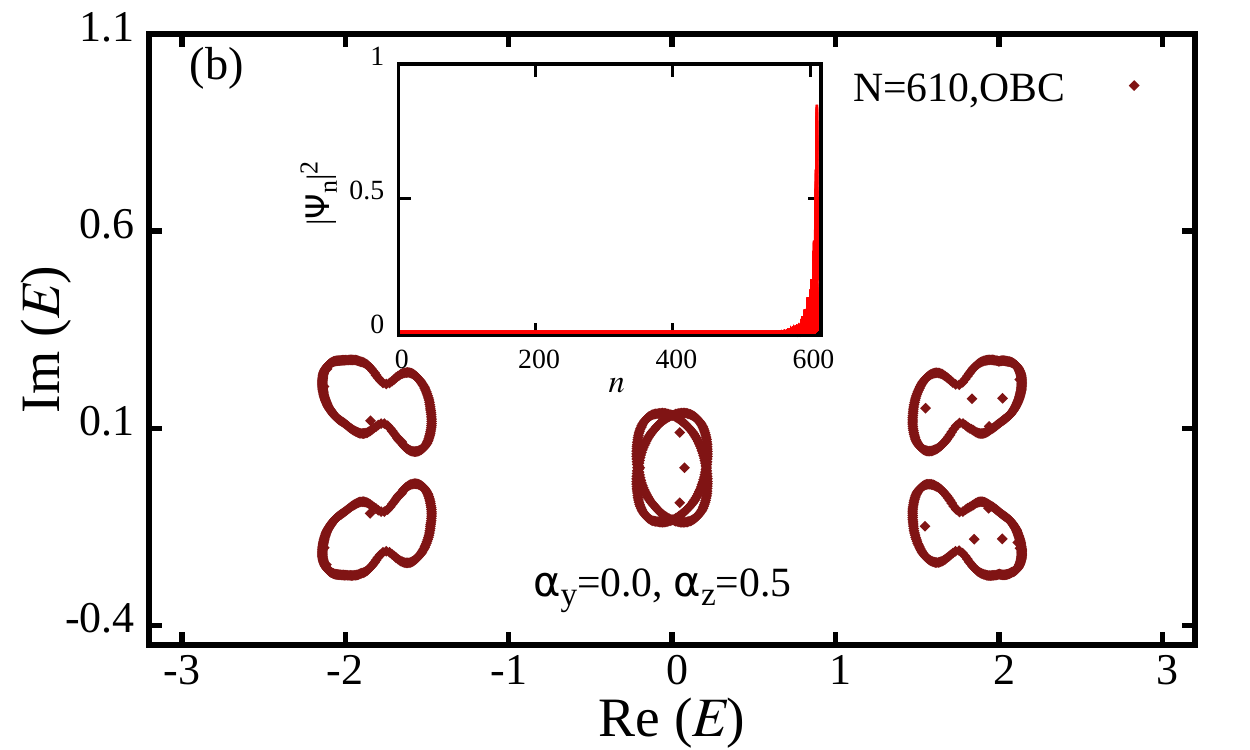}	
	    			\includegraphics[width=0.22\textwidth,height=0.20\textwidth]{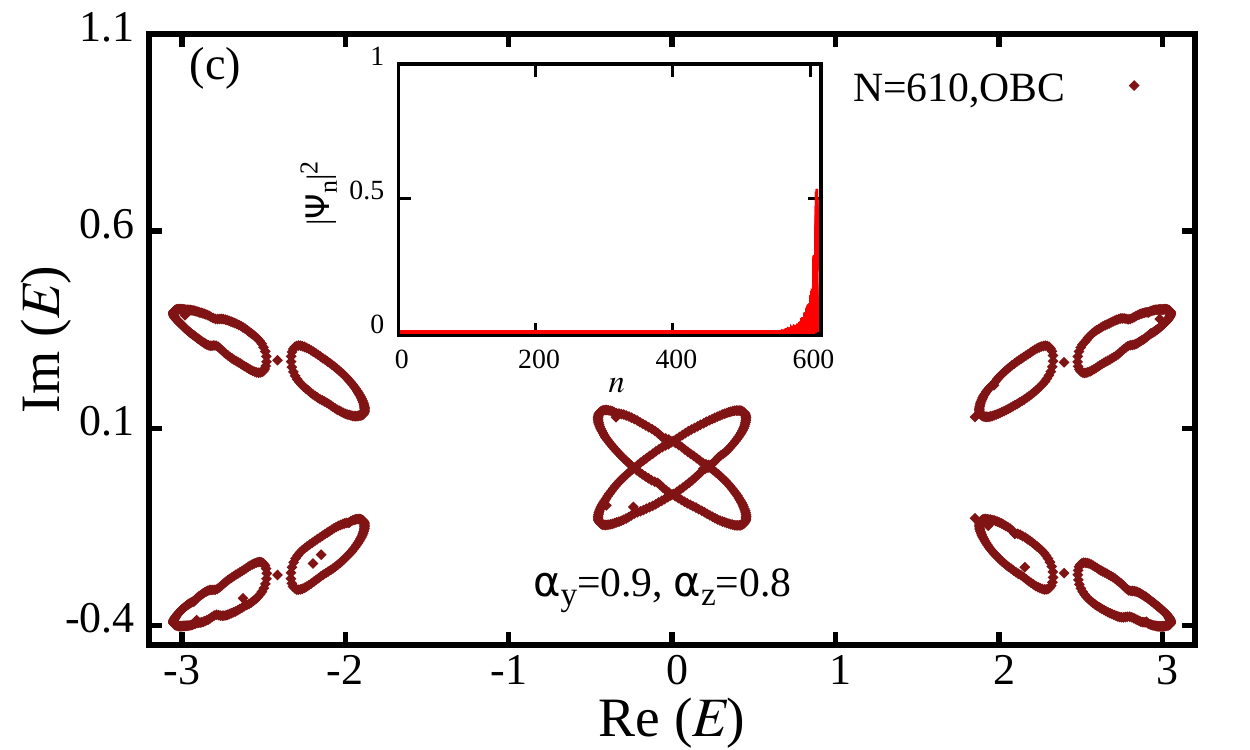} 	 
	    	\vspace{-0.2cm}
	    	\caption{The energy spectrum for the non-Hermitian Hamiltonian ($J_L/J_R=0.5$) defined in
	    	Eq.~(\ref{Eq:Total_Hamiltonian}) for different cases:
	        (a) PBC (blue) and OBC (black) in the absence of disorder ($V=0.0$) and without RSO interaction ($\alpha_y=0.0$ and $\alpha_z=0.0$). Here, N=55. (b) In the presence of only the spin-flip RSO interaction ($\alpha_y=0.5$ and $\alpha_z=0.0$).
	    	(c) In the presence of strong RSO interaction ($\alpha_y=0.9$ and $\alpha_z=0.8$). 
	    	In the last two figures, we have set $V=1.50$ for a 610-site system.
    		The density profile $|\psi_n|^2$ in all the three cases for OBC are shown in the insets.}
	    	\label{Fig:NH_skin_effect}
	    \end{figure}

    \section{BOUNDARY CONDITION DEPENDENCE AND SKIN EFFECT }\label{App:Skin_effect}
    
    The systems possessing non-reciprocity in the hopping amplitudes are dramatically sensitive to the choice of boundary conditions.
	Fig.~\ref{Fig:NH_skin_effect}(a) shows the complex-real transition in the energy spectrum for a non-reciprocal system with 55 sites.
	However, the energy spectrum no longer remains real on increasing the number of unit cells in the system, as demonstrated in Sec.~\ref{Sec:Skin_effect} of the main text. 
	Figs.~\ref{Fig:NH_skin_effect}(b-c) portrays the complex spectrum bearing the skin-effect in the presence of RSO interaction.
	It is evident from Fig.~\ref{Fig:NH_skin_effect}(b) and Fig.~\ref{Fig:Skin_effect}(c) that the interchange of $\alpha_y$ and $\alpha_z$ have no intuitive change in the skin-effect. 
	In Fig.~\ref{Fig:NH_skin_effect}(c), we have demonstrated the existence of the skin effect when both the RSO interaction strengths are non-zero.

	\begin{figure}
		\begin{center}	
			\begin{tabular}{p{\linewidth}c}
				\begin{center}
					
					\includegraphics[width=0.40\textwidth,height=0.25\textwidth]{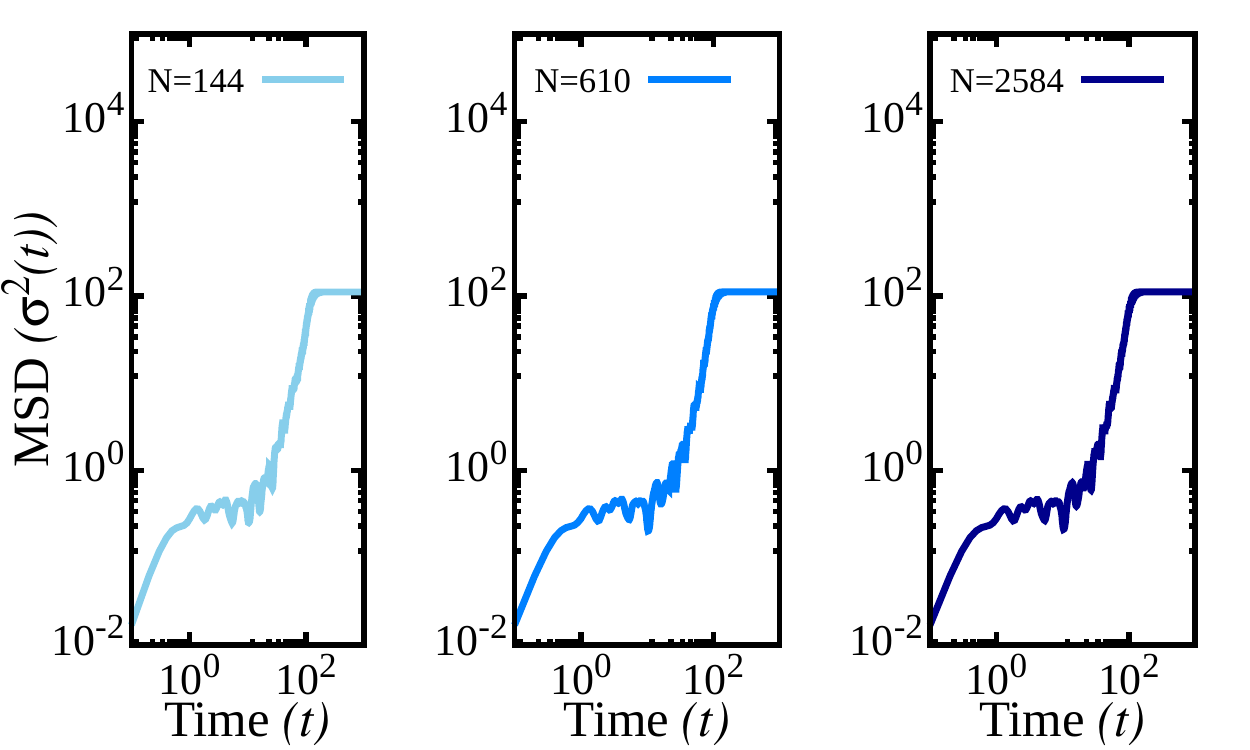}\hspace{0.0cm}
					
				\end{center}
			\end{tabular}	
		\end{center}
		\vspace{-0.8cm}
		\caption{The MSD for a non-Hermitian system ($J_L/J_R=0.5$) demonstrated for three different system sizes ($L$= 144, 610 and 2584) as illustrated by the 
		light blue, medium-blue and dark-blue lines respectively. Here, we have set $\alpha_y=0.5$, $\alpha_z=0.0$ and $V$=3.50 (spectrally localized regime) in 
		all the three cases.}
		\vspace{-0.5cm}
		\label{Fig:MSD_system_size_dependence}
	\end{figure}
	
	\section{ABSENCE OF FINITE SIZE AND BOUNDARY EFFECTS IN THE MSD RESULTS}\label{App:MSD_NH}
\begin{figure}
		\begin{center}	
			\begin{tabular}{p{\linewidth}c}
				\begin{center}
					
					\includegraphics[width=0.27\textwidth,height=0.25\textwidth]{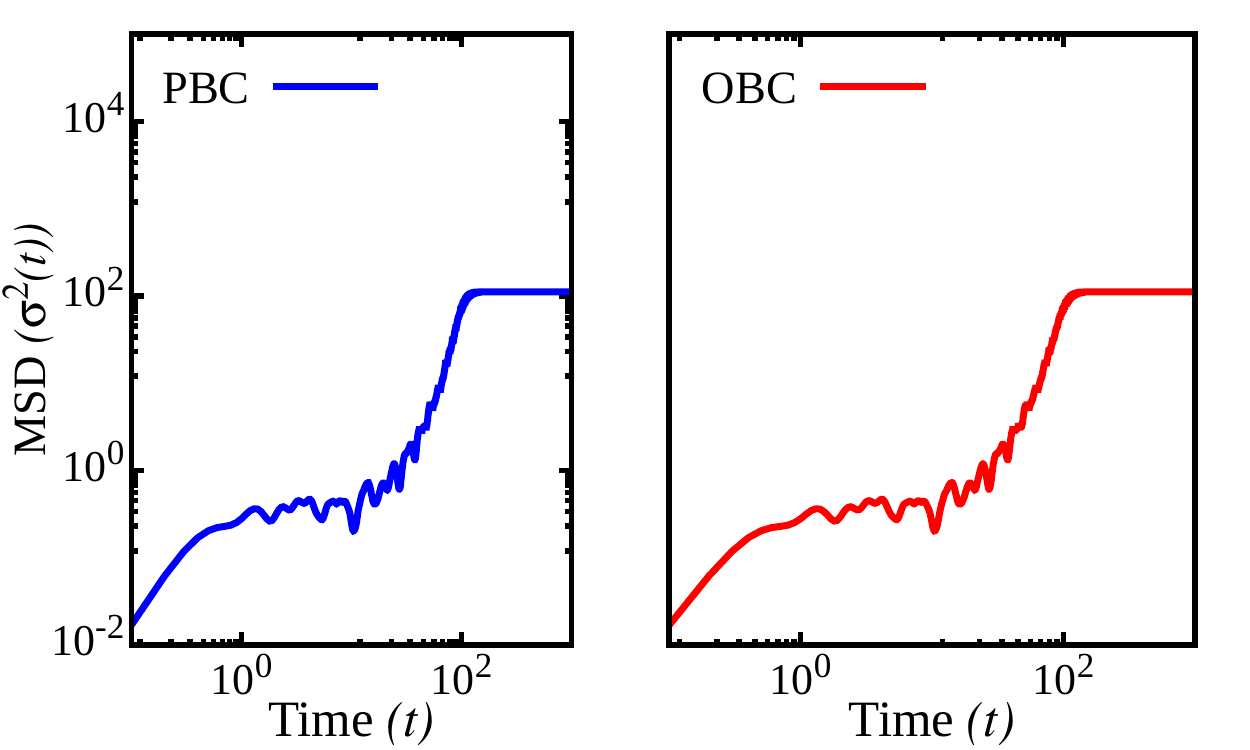}\hspace{0.0cm}
					
				\end{center}
			\end{tabular}	
		\end{center}
		\vspace{-0.8cm}
		\caption{The MSD for a non-Hermitian system ($J_L/J_R=0.5$, $\alpha_y=0.5$ and $\alpha_z=0.0$) under the periodic boundary (in blue) and open boundary (in red)
		conditions respectively for a lattice with $610$ sites. The quasiperiodic potential is set at $V$=3.50 (spectrally localized regime).}
		\vspace{-0.5cm}
		\label{Fig:MSD_boundary_dependence}
	\end{figure}

	Fig.~(\ref{Fig:MSD_system_size_dependence}) demonstrates the behavior of the MSD over asymptotically long times.
	We have verified that the numerical results are independent of the system size by using three different system sizes, i.e., $N=144,610$ and $2584$.
	However, all calculations in the main text have been estimated by considering a lattice of 610 sites.\\
	\indent
	In addition, as demonstrated in Appendix~\ref{App:Skin_effect}, the non-Hermitian system described by Eq.~(\ref{Eq:Total_Hamiltonian}) in the main text depends upon the choice of the boundaries.
	However, it is observed from Fig.~(\ref{Fig:MSD_boundary_dependence}) that the MSD estimates are independent of the boundary conditions.
	
	\begin{figure}
		\begin{center}	
			\begin{tabular}{p{\linewidth}c}
				\begin{center}
					
					\includegraphics[width=0.30\textwidth,height=0.25\textwidth]{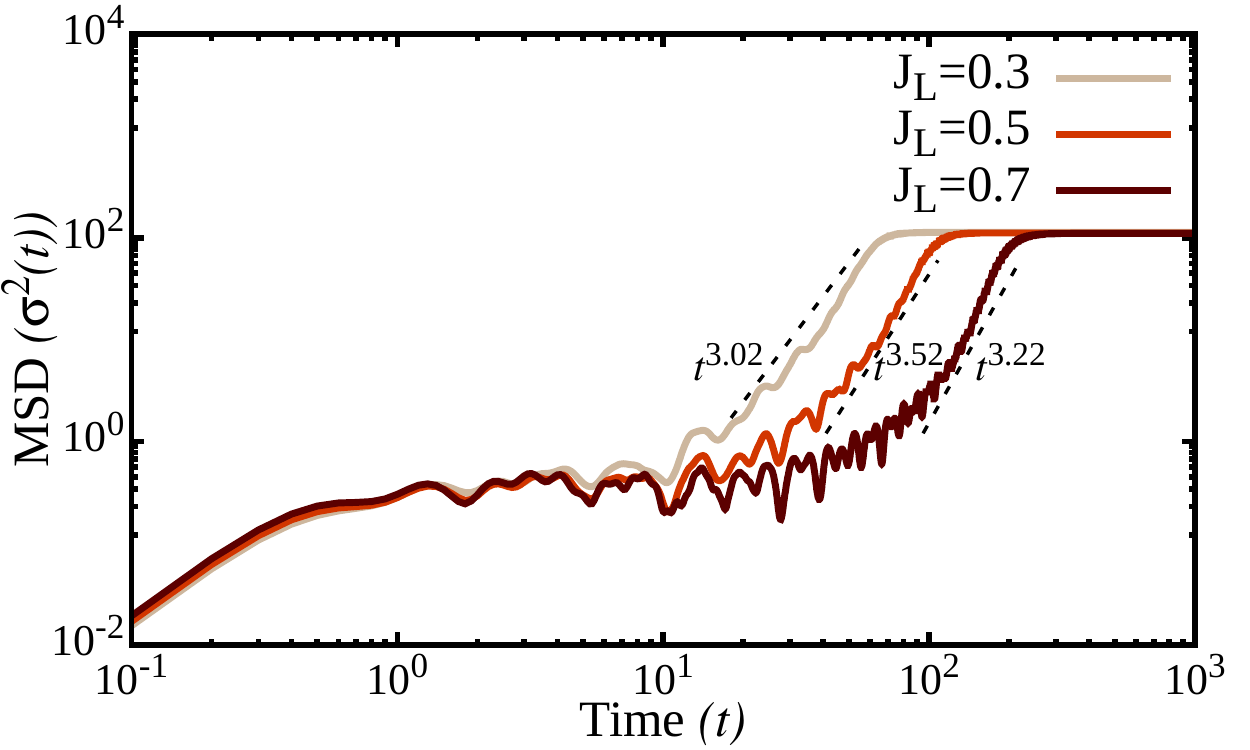}\hspace{0.0cm}
					
				\end{center}
			\end{tabular}	
		\end{center}
		\vspace{-0.9cm}
		\caption{The behavior of MSD as a function of time for different strengths of the non-Hermiticity in the spectrally
		localized ($V=3.50$) regime. Here, $J_R=1$, $\alpha_y=0.5$ and $\alpha_z=0.0$, and $N=610$ in all the three cases.}
		\label{Fig:MSD_NH_systems}
	\end{figure}

	\begin{figure}
		\begin{center}	
			\begin{tabular}{p{\linewidth}c}
				\begin{center}
					\centering
					\includegraphics[width=0.30\textwidth,height=0.25\textwidth]{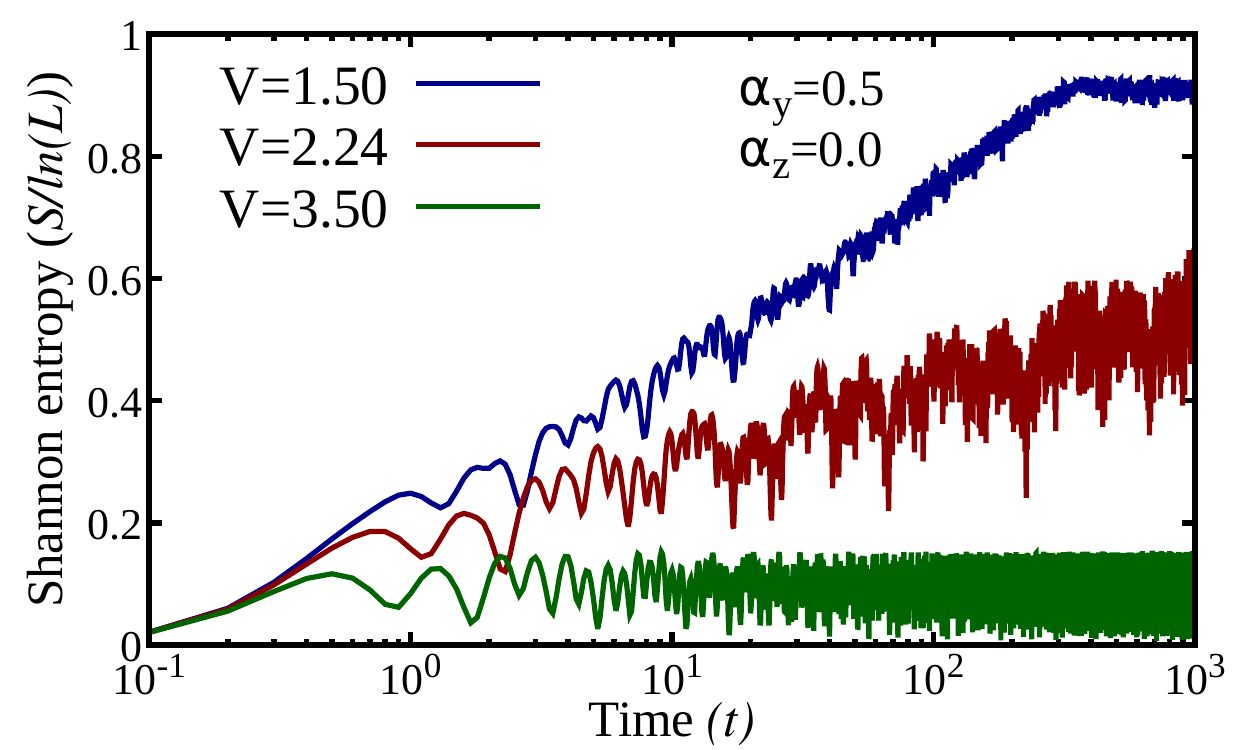}\hspace{0.0cm}	 
				\end{center}
			\end{tabular}	
		\end{center}
		\vspace{-0.8cm}
		\caption{The behavior of Shannon entropy for a Hermitian system ($J_L=J_R=1.0$) as a function of time. The dark-blue, dark-red and dark-green lines represents $S(t)$ in the delocalized, critical and localized regimes respectively.
			Here, $\alpha_y=0.5$ and $\alpha_z=0.0$ in all the cases. We have used periodic boundary condition on a lattice with 610 sites.}
		\label{Fig:Shannon_entropy_Hermitian}
	\end{figure}
		
	\section{COMPARISON OF MSD FOR DIFFERENT RATIOS OF \textbf{$J_L/J_R$}}\label{App:MSD}
	Fig.~(\ref{Fig:MSD_NH_systems}) illustrates the behavior of the MSD defined in Eq.~(\ref{Eq:Mean_Square_Displacement}) for different ratios of the non-reciprocal hopping amplitudes ($J_L/J_R=0.3,0.5~\text{and}~ 0.7$).
	It is a clear indicative that in the presence of RSO interaction, all the non-Hermitian systems exhibit dynamical delocalization (with hyper-diffusive behavior).
	However, the time for ultimate localization may vary depending upon the strength of the non-reciprocity.
   
	\section{SHANNON ENTROPY IN THE HERMITIAN SYSTEM WITH RSO}\label{App:Hermitian_Shannon_entropy} 
	Fig.~(\ref{Fig:Shannon_entropy_Hermitian}) illustrates the behavior of Shannon entropy with time. 
	As described in Sec.~\ref{Sec:Shannon_entropy} of the main text, at long times, the entropy increases due to the itinerant behavior of the electrons and $S/ln L$ approaches 1 in the extended regime as shown by the dark-blue line in Fig.~(\ref{Fig:Shannon_entropy_Hermitian}).
	In the localized regime, the entropy is diminished and remains constant over time (shown by dark-green line).
	The behavior in the critical regime lies intermediate bewteen the extended and localized states (dark-red line).

\end{document}